\documentclass[journal]{IEEEtran}

\usepackage{cite}
\usepackage{amssymb}
\usepackage{amstext}
\usepackage{amsmath}
\usepackage{amsthm}
\usepackage{multicol}
\usepackage{float}
\usepackage{hyperref}
\usepackage{subfig}

\usepackage{xcolor}
\ifCLASSINFOpdf
  \usepackage[pdftex]{graphicx}
\else  
  \usepackage[dvips]{graphicx}
\fi

\usepackage{url}
\usepackage{listings}
\usepackage[mathscr]{eucal}
\def\postbreak{\raisebox{0ex}[0ex][0ex]{\ensuremath{\hookrightarrow\space}}}
\def\sign{\mbox{sign\,}}

\newcommand{\br}{\mathbb{R}}
\newcommand{\oms}{\text{\scriptsize $\mathscr{O}$}}

\newtheorem{lemma}{Lemma}
\newtheorem{theorem}{Theorem}

\newtheorem{definition}{Definition}
\newtheorem{ass}{Assumption}
\newtheorem{prop}{Proposition}

\begin{document}
\title{Nonlinear analysis of charge-pump phase-locked loop: 
the hold-in and pull-in ranges}
\author{
    Kuznetsov~N.V.$^{\,a,b,c}$, Matveev~A.S.$^{\,a}$, Yuldashev~M.V.$^{\,a}$, Yuldashev~R.V.$^{\,a}$\\
\thanks{
($^a$) Faculty of Mathematics and Mechanics,
Saint-Petersburg State University, Russia;
($^b$) Faculty of Information Technology,
University of Jyv\"{a}skyl\"{a}, Finland;
($^c$) 
Institute for Problems in Mechanical Engineering of Russian Academy of Science, Russia;
(corresponding author email: n.v.kuznetsov@spbu.ru, nkuznetsov239@gmail.com).
}
}

\markboth{IEEE Transactions on Circuits and Systems I: Regular Papers}%
{Submitted paper}

 \IEEEaftertitletext{
\begin{flushright} 
 {\it Dedicated to the memory of IEEE fellow Floyd M. Gardner (1929 – 2021).}
\end{flushright}
 }
\maketitle

\begin{abstract}
In this paper a fairly complete mathematical model of CP-PLL,
which reliable enough to serve as a tool for credible analysis of
dynamical properties of these circuits, is studied. 
We refine relevant mathematical definitions of 
the hold-in and pull-in ranges related to the local and global stability.
Stability analysis of the steady state for the charge-pump phase locked loop is non-trivial: 
straight-forward linearization of available CP-PLL models
may lead to incorrect conclusions, 
because the system is not smooth near the steady state and 
may experience overload. 
In this work necessary details for local stability analysis are presented
and the hold-in range is computed.
An upper estimate of the pull-in range is obtained 
via the analysis of limit cycles.
The study provided an answer to Gardner’s conjecture on the similarity of transient responses of CP-PLL and equivalent classical PLL
and to conjectures on the infinite pull-in range of CP-PLL 
with proportionally-integrating filter.
\end{abstract}


\IEEEpeerreviewmaketitle

\section{Introduction}

Design and analysis of frequency control circuits
is a challenging task relevant to many applications:
satellite navigation \cite{KaplanH-2006-GPS},
digital communication \cite{Proakis-2007},
wireless networks \cite{DuS-2010-communication},
to mention just a few. Effective locking onto the phase of the input signal is among the
principal problems solved by means of such circuits. From a broad perspective,
their synthesis and analysis fall under the framework of standard topics in control
engineering like signal tracking, linear and global stability.
Meanwhile, some of ubiquitous and actively used circuits are largely
inspired by implementability issues and approaches of practical control
engineering so that their true capacities and limitations still await fully
disclosing via a rigorous analysis.
\par
This paper aims at filling this gap with respect to the
Charge-Pump Phase-Locked Loop (CP-PLL),
which is used for frequency synthesis and clock generation in computer architectures \cite{Bianchi-2005-book}.
The CP-PLL is able to quickly lock onto the phase of the incoming signal,
achieving low steady-state phase error.
Stability of the CP-PLL steady state (the locked state) was originally studied 
by F.~Gardner in \cite{Gardner-1980} using approximate linear models.
In this pioneering work he conjectured that {\it transient response of practical charge-pump PLL's can be expected to be nearly the same as the response of the equivalent classical PLL}. 
Later on, approximate discrete-time linear models of the CP-PLL were suggested in \cite{Hein-1988,lu2001discrete}.
The closed loop nonlinear discrete time model of CP-PLL was suggested in 
\cite{Paemel-1994}.
In this paper we develop, augment, and supplement the approach used in the reported literature in order to extend it to the practically important case of Voltage Controlled Oscillator (VCO) overload 
(see, e.g. \cite{GillespieKK-2000,KuznetsovYYBKM-2019-arXiv}).

The range of input frequencies associated with stable steady state corresponds to the hold-in range.
For the classical analog PLL, stability of the locked state depends on the gap between the VCO free-running frequency and the frequency of the reference signal.
For first-order active proportionally-integrating (PI) filter, analog PLL is theoretically stable for any gap \cite{KuznetsovLYY-2021-TCASII}.
Conversely, stability of the locked state of CP-PLL depends on the reference frequency even if PI loop filter is employed.
Moreover, the CP-PLL is stable only for relatively high input frequencies, 
which is far different from that with stability of analog PLLs. 
It follows that even the definitions of the hold-in, the pull-in and the lock-in ranges (see, e.g. \cite{KuznetsovLYY-2015-IFAC-Ranges,LeonovKYY-2015-TCAS,BestKLYY-2016,KuznetsovLYY-2019-DAN}) should be refined for the CP-PLL, to say nothing about the need to update and extend the base of relevant knowledge about the properties of the circuit.
Note, that it is important to know how VCO overload limits stability, because CP-PLL may not operate correctly near the overload due to non-linearities of voltage-frequency characteristics.
From practical point of view overload should be avoided, 
since it may break circuits or lead to unpredicted behaviour \cite{KuznetsovMYYB-2020-IFACWC}.
Therefore it is important to find parameters which may lead to overload and understand how to identify oscillations tied to it.

Extra troubles stem from the fact that straight-forward linearization of available CP-PLL models may lead to incorrect conclusions, because the system is not smooth near the steady state (in fact, it is only piecewise smooth).
In \cite{Orla-2013-review}, stability analysis follows the lines of a Lyapunov approach, however, details of the proof are not presented.
Note that nonlinear high-order mathematical models of CP-PLL
can also be built by using approximations of exponentials
(see, e.g. \cite{Hedayat-1999-3order,Johnson-2005,Wang-2005,Daniels-2008,Guermandi-2011,bizzarri2012periodic,Shakhtarin2014-pfd,Hedayat-2014-high-order}), but the resulting transcendental equations can not be solved analytically without using approximations.

In this paper, we use the findings of \cite{KuznetsovYYBKKM-2019} as a keystone, and develop, augment, and supplement them in order to acquire 
a fairly complete mathematical model of CP-PLL reliable enough to serve as a tool for credible analysis of dynamical properties of these circuits. To this end, we also refine some relevant mathematical definitions of main characteristics, and demonstrate the potentiality of the proposed model.

\section{Mathematical model of the charge-pump phase-locked loop with
phase-frequency detector}
\label{sec:math-model}
\begin{figure*}[!ht]
  \centering
  \includegraphics[width=0.95\linewidth]{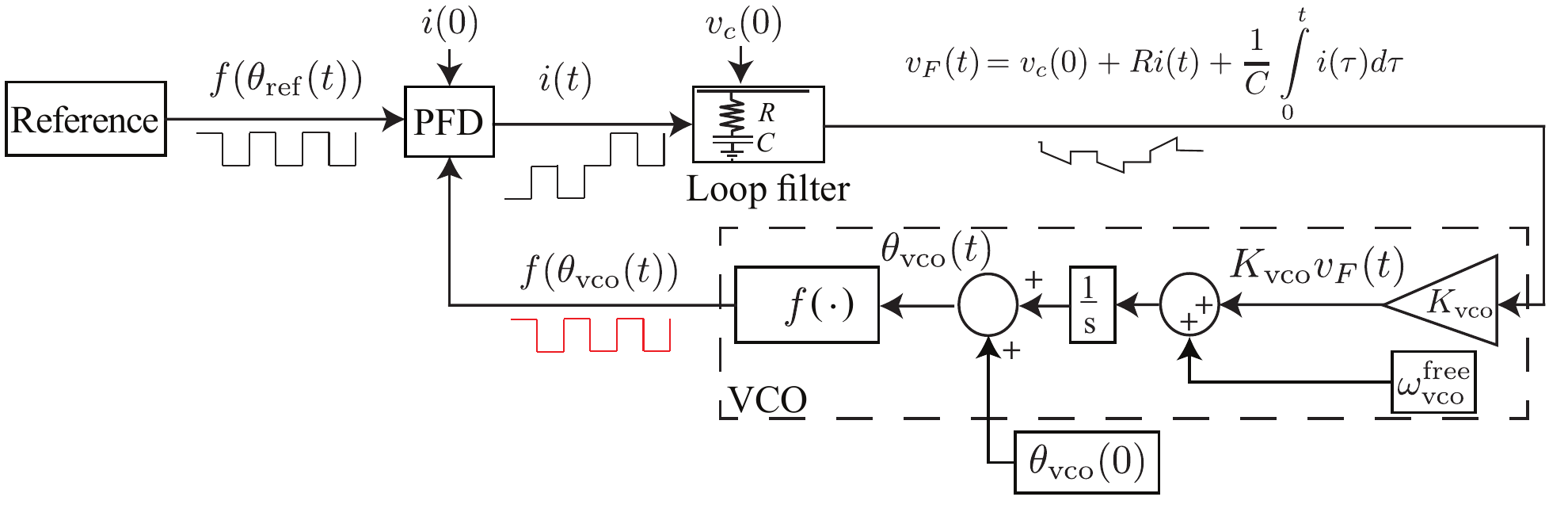}
  \caption{Charge-pump PLL.}
  \label{pfd_char}
\end{figure*}

\begin{figure}[!ht]
  \centering
  \includegraphics[width=0.5\linewidth]{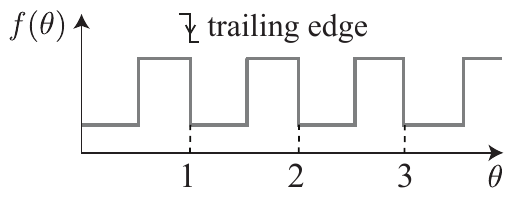}
  \caption{Waveform of the reference and VCO signals.}
\label{f1-f2}
\end{figure}

Consider the charge-pump phase-locked loop with
phase-frequency detector \cite{Gardner-1980,Gardner-2005-book} in Fig.~\ref{pfd_char}.
Both the reference (Ref) and output of the VCO are square waveform signals (Fig.~\ref{f1-f2}) with phases $\theta_{\rm vco}(t)$ and $\theta_{\rm ref}(t)$, respectively.
Without loss of generality we suppose that
trailing edges of the VCO and reference signals occur when the corresponding phase reaches an integer number.
The frequency $\omega_{\rm ref}>0$ of reference signal (reference frequency)
is usually assumed to be constant:
\begin{equation}\label{omega-ref}
  \theta_{\rm ref}(t) = \omega_{\rm ref}t = \frac{t}{T_{\rm ref}},
\end{equation}
where $T_{\rm ref}>0$ 
is a period
of the reference signal.

\begin{figure}[H]
\centering
  \includegraphics[width=0.95\linewidth]{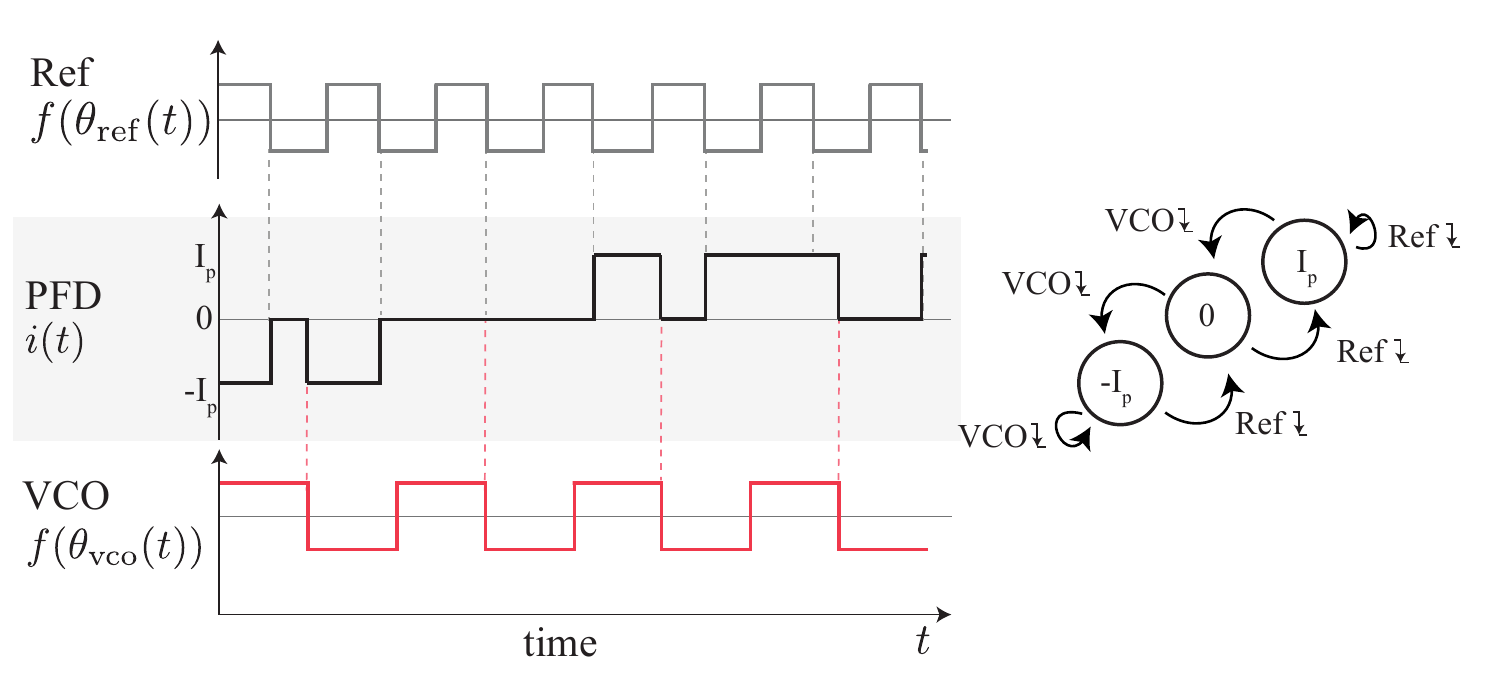}
    \caption{Phase-frequency detector operation.}
  \label{pfd}
\end{figure}

The Phase-Frequency Detector (PFD) is a digital circuit, triggered by the trailing (falling) edges of the Ref and VCO signals. The output signal of PFD $i(t)$
can have only three states (Fig.~\ref{pfd}): 0, $+I_p$, and $-I_p$.
A trailing edge of the reference signal forces the PFD to switch to a higher state,
unless it is already in the state $+I_p$.
A trailing edge of the VCO signal forces the PFD to switch to a lower state,
unless it is already in the state $-I_p$.
If both trailing edges happen at the same time, then the PFD switches to zero.
Therefore, for positive frequencies the PFD output $i(t)$ always returns to zero
from non-zero state at certain time.
Thus, without loss of generality, assume that $i(t)=0$ till the time $t=0$ 
(i.e. $i(0-)=i(0)=0$)
when for the first time one of trailing edge appears at the input of the PFD.
For $t>0$ the function $i(t)$ is determined by $i(t-)$, $\theta_{\rm vco}(t)$, and $\theta_{\rm ref}(t)$.

The relationship between the input current $i(t)$
and the output voltage $v_F(t)$ for a
proportionally integrating (perfect PI) filter
based on  resistor and capacitor
is as follows
\begin{equation}
\label{RC-filter}
  v_F(t) = v_c(0) + Ri(t) + \frac{1}{C}\int\limits_0^t i(\tau)d\tau,
\end{equation}
where $R>0$ is a resistance, $C>0$ is a capacitance,
and $v_c(t)=v_c(0) + \tfrac{1}{C}\int\limits_0^t i(\tau)d\tau$ is a capacitor charge; the transfer function is $H(s) = R + \frac{1}{Cs}$.

The control signal $v_F(t)$ adjusts the VCO frequency:
\begin{equation} \label{vco first}
        \dot\theta_{\rm vco}(t) =
         \omega_{\rm vco}(t) =
         \omega_{\rm vco}^{\text{free}} + K_{\rm vco}v_F(t),
\end{equation}
where $\omega_{\rm vco}^{\text{free}}$ is the VCO free-running (quiescent) frequency
(i.e. for $v_F(t)\equiv 0$), $K_{\rm vco}$
is the VCO gain (sensitivity). 

\begin{figure}[H]
\centering
\includegraphics[width=0.4\linewidth]{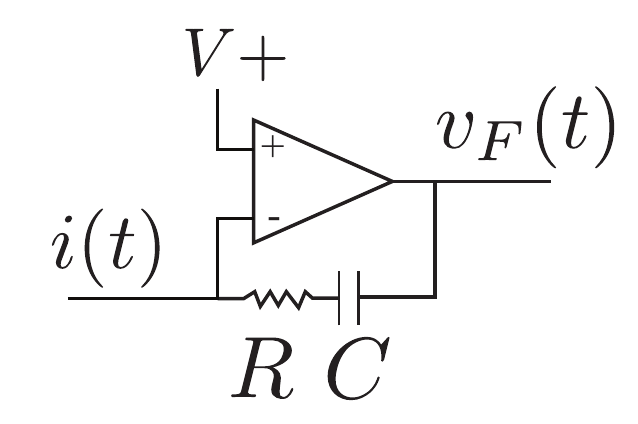}
\caption{Loop filter with operational amplifier:
the transfer function $H(s) = I_p(R + (Cs)^{-1})$.}
\label{fig:cppll-opamp-filter}
\end{figure}

Consider one important thing regarding the charge-pump (CP) using in the PFD:
transistors inside CP reasonably approximates the current generators until the drain-source voltage magnitude is higher than a given minimum value.
Note that both transistors’ output characteristics approximate the current generators only if the output voltage is within the current saturation region.
The CP-PLL will work as expected only if the CP output voltage is within its valid range.
In order to keep
both transistors within their current saturation region, a ``zero impedance'' 
(or very low) to ground (or to any DC voltage) is needed. 
This means that the loop
filter has to have a capacitor to GND if it is purely passive.
One of the simplest solutions is to use a second-order filter.
Here we consider another solution: to add operational amplifier to the filter \cite{KuznetsovMYYB-2020-IFACWC}
(see Fig.~\ref{fig:cppll-opamp-filter}).
In this case the transfer function of the loop filter remains the same as in \eqref{RC-filter}.

From \eqref{omega-ref}, \eqref{RC-filter}, and \eqref{vco first},
for the given 
$\omega_{\rm ref}$
we obtain a \emph{continuous time nonlinear mathematical model of CP-PLL}
described by the following differential equations
\begin{equation}\label{CPPLL-signal-space-model}
\begin{aligned}
  & \dot v_c(t) = \tfrac{1}{C}i(t),
  \
   \dot\theta_{\rm vco}(t)  =
        \omega_{\rm vco}^{\text{free}}\!+\!K_{\rm vco}
        \left(
          Ri(t)\!+\!v_c(t)
        \right)
\end{aligned}
\end{equation}
with the discontinuous, piecewise constant nonlinearity
\(
  i(t) = i\big(i(t-), \omega_{\rm ref}, \theta_{\rm vco}(t)\big)
\)
and initial conditions $\big(v_c(0), \theta_{\rm vco}(0)\big)$.
This model is   nonlinear, non-autonomous, discontinuous, and switching system which is hard to analyze.

\subsection{Overload}
Depending on the design, the VCO input in CP-PLL 
may experience overload 
(see, e.g. \cite{Paemel-1994,KuznetsovMYYB-2020-IFACWC,KuznetsovYYBKM-2019-arXiv}).
From the mathematical point of view the VCO overload means that frequency is driven to a non-positive value, i.e at some point $t'$
\begin{equation}\label{omegavcooverload}
   \dot\theta_{\rm vco}(t') = \omega_{\rm vco}^{\text{free}}\!+\!K_{\rm vco}
        \left(
          Ri(t')\!+\!v_c(t')
        \right)\leq 0.
\end{equation}

\subsection{Locked states}

If the synchronization is achieved, i.e. a transient process is over,
then the loop is said to be in a \emph{locked state}.
When CP-PLL is in a locked state,
the trailing edges of the VCO signal happen almost at the same time as the trailing edges of the reference signal. 
In a locked state the output of PFD $i(t)$
can be non-zero
only on short time intervals (shorter than $\tau_{\rm lock}$).
The allowed residual phase difference $\tau_{\rm lock}$
should be in agreement with engineering requirements
for a particular application. 
We consider the ideal case $\tau_{\rm lock}=0$.

\section{Nonlinear discrete time CP-PLL model}
\begin{figure}[H]
\centering
  \includegraphics[width=0.95\linewidth]{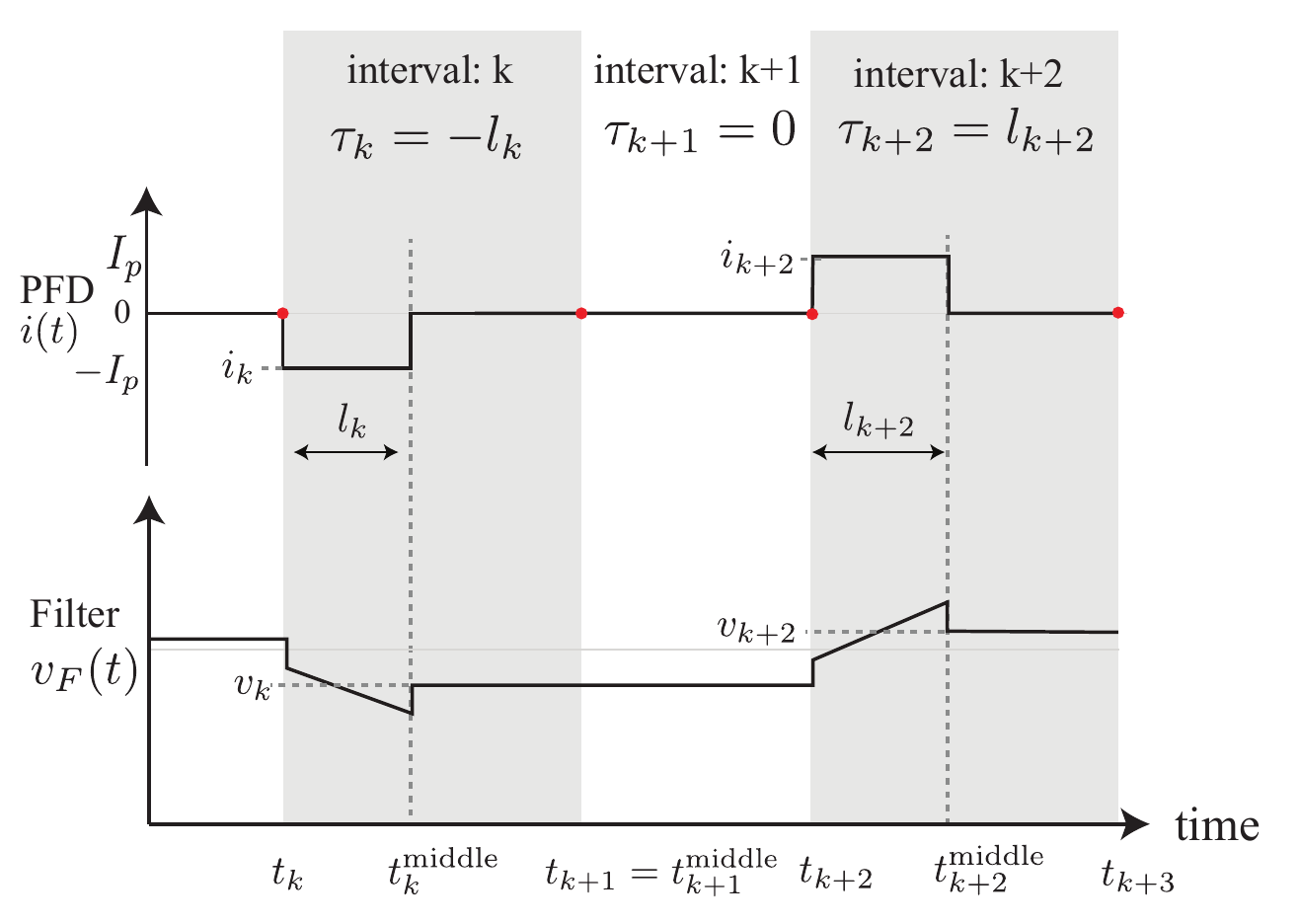}
  \caption{Discrete states $\tau_k$ and $v_k$; $l_k$ is the PFD pulse width.}
  \label{fig states tau v}
\end{figure}

For nonlinear analysis, we pass from model \eqref{CPPLL-signal-space-model} to a discrete-time model.
Following \cite{KuznetsovYYBKKM-2019}, we derive a discrete time model of the CP-PLL. 
Without loss of generality, assume that $i(t)=0$ till the time $t_0$ 
(i.e. $i(t_0)= 0$)
when for the first time one of trailing edge appears at the input of the PFD.
If only one of the trailing edges appear, the PFD output becomes non-zero 
(i.e. $i(t_0+)\neq 0$),
and then we denote by $t_0^{\rm middle}> t_0$ the first instant of time such that the PFD output becomes zero (i.e. $i(t_0^{\rm middle})= 0$).
If both VCO and Ref trailing edges appear at the same time, we denote $t_0^{\rm middle}=t_0$.
Then, we wait until the first trailing edge of the VCO or Ref,
and denote the corresponding moment of time by $t_1$ (i.e. $i(t_1)=0$). 
Continuing in a similar way, one obtains
the increasing sequences $\{t_k\}$ and $\{t_k^{\rm middle}\}$ for $k=0,1,2...$
Thus, $i(t)=0$ for $t_k^{\rm middle}\leq t \leq t_{k+1}$ and $i(t)=\pm I_p$ for $t_k<t< t_k^{\rm middle}$.
Denote by $\tau_k$ the PFD pulse width (length of the time interval,
where the PFD output is a non-zero constant)
multiplied by the sign of the PFD output (see Fig.~\ref{fig states tau v}):
\begin{equation}
\label{tau k def}
\begin{aligned}
  \tau_k = 
  \left(t_k^{\rm middle} - t_k\right)
  \sign\left( i\left(\frac{t_k+t_k^{\rm middle}}{2}\right)\right)
\end{aligned}
\end{equation}

If the VCO trailing edge hits before the Ref trailing edge
then $\tau_k < 0$, in the opposite case we have $\tau_k > 0$.
Thus, $\tau_k$ shows how one signal lags behind another.

From \eqref{RC-filter} it follows that
the zero output of PFD $i(t) \equiv 0$ on the interval $[t_k^{\rm middle},t_{k+1}]$
implies a constant filter output.
Denote this constant by $v_k$. We have
\begin{equation}
\label{v_k def}
\begin{aligned}
  & v_F(t) \equiv v_k, \quad t \in [t_k^{\rm middle},t_{k+1}].
\end{aligned}
\end{equation}
Following the ideas from \cite{acco2003etude,Orla-2013-review},
denote
\begin{equation}
\label{eq:alpha-beta-eqs}
  \begin{aligned}
    &  p_k = \frac{\tau_k}{T_{\rm ref}},\
      u_k=T_{\rm ref}
      \left(
        \omega_{\rm vco}^{\text{free}} + K_{\rm vco}v_k
      \right) - 1,
      \\
    &
      \alpha = K_{\rm vco}I_pT_{\rm ref}R>0,\
     \beta = \frac{K_{\rm vco}I_pT_{\rm ref}^2}{2C}>0.
  \end{aligned}
\end{equation}
Here $p_k$ is a normalized phase shift and $u_k+1$ is a ratio of the VCO frequency $\omega_{\rm vco}^{\text{free}} + K_{\rm vco}v_k$ to the Ref frequency $\frac{1}{T_{\rm ref}}$.
\subsection{Overload}
  Condition~\eqref{omegavcooverload} of the VCO overload 
  on $[t_k,t_{k+1})$
  corresponds to the following cases:
\begin{itemize}
  \item for $\tau_k<0$ one can see from Fig.~\ref{fig:vco-overload-alpha-1} that the VCO is overloaded for 
  \begin{equation}
    \label{app:eq:vco-overload-cond-alpha}
      {\inf\limits_{t\in[t_k,t_{k+1})}\!\!\!\!\!\!\!\dot\theta_{\rm vco}(t)}
      \!=\!
      \omega_{\rm vco}^{\text{free}}\!+\!K_{\rm vco}(v_c(t_k)\!+\!\tau_k\tfrac{I_p}{C}-I_p R)\!\leq\!0;
  \end{equation}
  \begin{figure}[ht]
    \centering
    \includegraphics[width=0.4\linewidth]{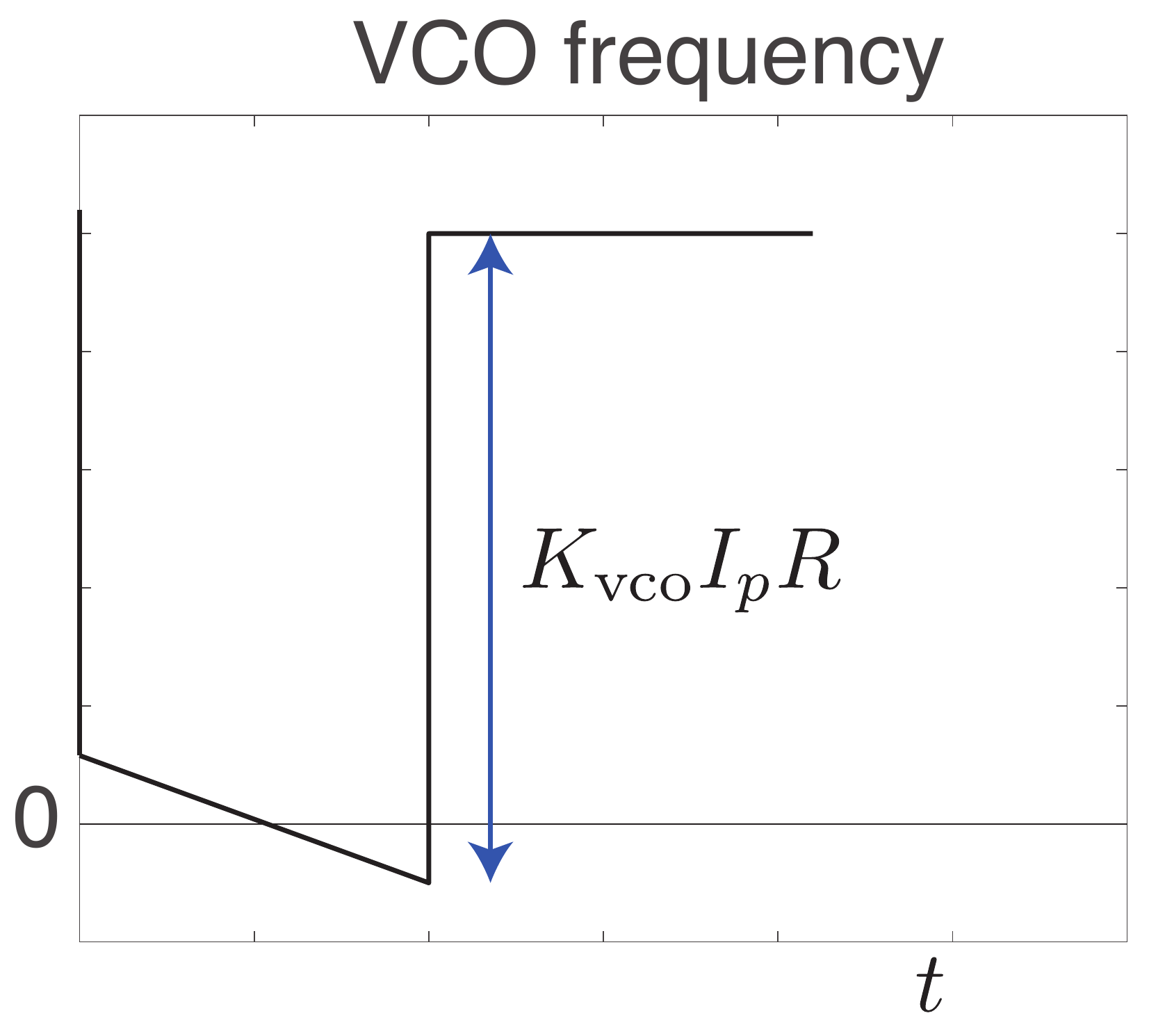}
    \includegraphics[width=0.4\linewidth]{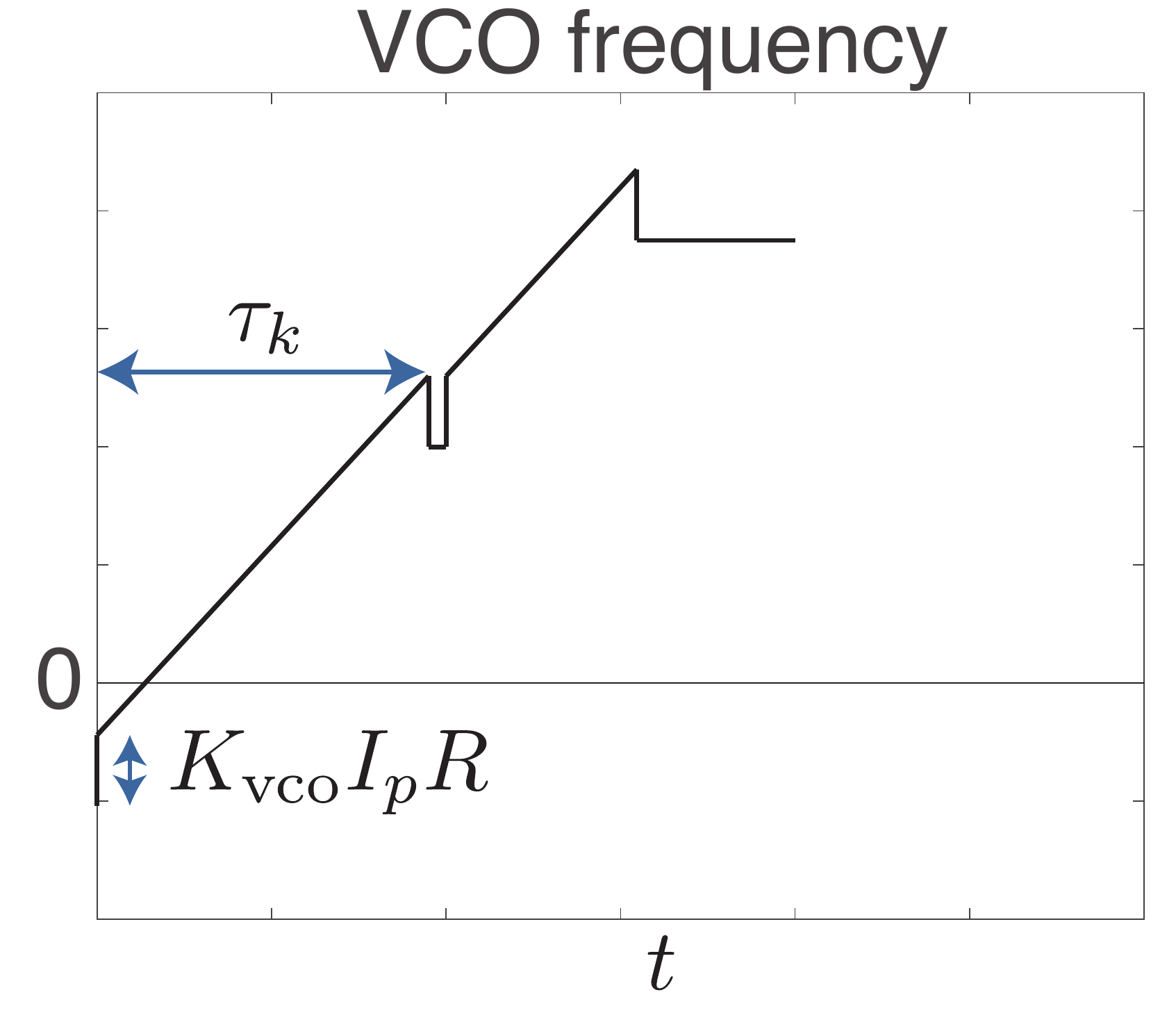}
    \caption{VCO overload near the locked state: a) for $\tau_k<0$
     b) for $0<\tau_k<1$
    }
    \label{fig:vco-overload-alpha-1}
  \end{figure}
\item 
for $\tau_k \geq 0$, the VCO may be also overloaded initially, see Fig.~\ref{fig:vco-overload-alpha-1}:
  \begin{equation}
    \label{app:eq:vco-overload-cond-beta}
      {\inf\limits_{t\in[t_k,t_{k+1})}\dot\theta_{\rm vco}(t)}=\omega_{\rm vco}^{\text{free}}+K_{\rm vco}v_c(t_k)\leq0.
  \end{equation}
\end{itemize}

In practice the VCO overload should be avoided (see, e.g. \cite{Paemel-1994,KuznetsovMYYB-2020-IFACWC}).
From the mathematical point of view a task may be posed 
to find the biggest positively invariable region of phase space 
in which there is no overload.
However, for any parameters $R$, $C$, $K_{\rm vco}$, and $I_p$ the VCO overload may occur 
for sufficiently large frequency difference between the VCO and reference signals.
Therefore, it is reasonable to demand that at least 
for the same frequencies of the Ref and VCO signals
and small time delay between signals
there is no overload.
For locked state, i.e. for $\tau_k=0$ and the same frequencies, 
inequality \eqref{app:eq:vco-overload-cond-beta} 
cannot be valid for $T_{\rm ref}>0$.
Thus, theoretically, in a locked state the VCO is not overloaded.
However the VCO can be overloaded near the locked state (\emph{local overload})
for any small width pulses 
on the output of the PFD.
Therefore, both equations \eqref{app:eq:vco-overload-cond-alpha} and \eqref{app:eq:vco-overload-cond-beta} should be checked.
Substituting $\frac{1}{T_{\rm ref}}=\omega_{\rm vco}^{\text{free}}+K_{\rm vco}v_c(t_k)$ into \eqref{app:eq:vco-overload-cond-alpha} and \eqref{app:eq:vco-overload-cond-beta} for $\tau_k \to 0$ we get

\begin{equation}
  \label{eq:local-overload}
  \begin{aligned}
    & \frac{1}{T_{\rm ref}}-K_{\rm vco}I_p R \leq 0,\text{ i.e. }  
    \alpha \geq 1.
  \end{aligned}
\end{equation}
Therefore, under following condition on the period of the input signal:
    $0<T_{\rm ref} < T^{\rm local}_{\rm overload}=
        \frac{1}{K_{\rm vco}I_pR}$, there is no local overload.
It may also be necessary to avoid the VCO overload even if 
the initial frequencies are equal but there 
is large time delay between signals, 
e.g. at startup or if the time delay between signals is changed due to a noise or period jump of the reference signal (\emph{startup overload}).
Using equal initial VCO and reference frequencies $\omega_{\rm vco}^{\text{free}}+K_{\rm vco}v_c(0)=\frac{1}{T_{\rm ref}}$ 
in combination with \eqref{app:eq:vco-overload-cond-alpha} and \eqref{app:eq:vco-overload-cond-beta} and assuming the worst case $\tau_1=-T_{\rm ref}$
we get startup overload condition
\begin{equation}
\label{eq:1}
\begin{aligned}
  & 1\!-\!\tfrac{K_{\rm vco}I_p T_{\rm ref}^2}{C}\!-\!K_{\rm vco}I_pT_{\rm ref}R
  \leq0,\text{ i.e. }
  1\!-\!2\beta\!-\!\alpha\!\leq\!0.
  \end{aligned}
\end{equation}
Thus, to avoid startup overload one should choose
\begin{equation}
  \begin{aligned}
  & 0<T_{\rm ref} 
  <T_{\rm overload}^{\rm startup}
  =\frac{-RC+\sqrt{(RC)^2+4\frac{C}{K_{\rm vco}I_p}}}{2},\\
    &\text{ i.e. }
    \alpha<1-2\beta.
  \end{aligned}
\end{equation}

\subsection{CP-PLL discrete time equations}

Final system of equations: 
$
x_{k+1}=f(x_k),
x_k = (u_k;p_k)
$,
describing CP-PLL outside of
the VCO overload
is obtained from \eqref{CPPLL-signal-space-model}, 
\eqref{tau k def}---\eqref{eq:alpha-beta-eqs} as follows
(see \cite{KuznetsovYYBKM-2019-arXiv} for details)\footnote{The 
differences between model \eqref{eq:complete-model-ab}  
and the models from \cite{Orla-2013-review} and \cite{Paemel-1994} can be demonstrated, for example, as follows.
 The values $\alpha = 0.2$, $\beta = 1.7$, $p_0 = -0.18$ and $u_0 = -0.43$
 are not taken into account in \cite{Orla-2013-review} where
 the corresponding values $a=\alpha$, 
 $b=\beta$, $\hat\tau_k=p_k$, and $v_k=u_k$ are outside of the allowed area 
 in Fig.~2.a \cite[page 1114]{Orla-2013-review}.
 The  values $R=0.2$, $C=0.01$, $K_{\rm vco}=20$, $I_p=0.1$, $T_{\rm ref}=0.125$, $\tau_0=0.0125$, and $v_0=1$ correspond to the VCO overload,
 while in \cite{Paemel-1994} the consideration of the corresponding values 
$R_2=0.2$, $C=0.01$, $K_v=20$, $I_p=0.1$, $T=0.125$, $\tau(0)=0.0125$, $v(0)=1$ leads to extracting of the square root from $-1$, i.e. $\sqrt{-1}$
(the editors of IEEE TCOM journal, where two pioneering works
\cite{Gardner-1980} and \cite{Paemel-1994} on the CP-PLL had been published, 
were notified about the problem and above results but did not provide a way to inform journal's readers). See details in \cite{KuznetsovYYBKKM-2019}.
}
\begin{equation}
\label{eq:complete-model-ab}
  \begin{aligned}
    & u_{k+1} =   u_k +2\beta p_{k+1},\\
    &  p_{k+1}=
    \begin{cases}
      \frac{-(u_k + \alpha + 1) + \sqrt{(u_k + \alpha + 1)^2 - 4\beta c_k}}{2\beta},
      \\ 
      \qquad\qquad \text{ for }  p_k \geq 0, \quad c_k \leq 0, 
      \\
      \frac{1}{ u_k + 1} -1 + ( p_k\text{ mod }1),
      \\ 
      \qquad\qquad \text{ for }  p_k \geq 0, \quad c_k > 0, 
      \\
      l_k-1,
      \\ 
      \qquad\qquad \text{ for } p_k < 0, \quad l_k \leq 1, 
      \\
      \frac{-(u_k + \alpha + 1) + \sqrt{(u_k + \alpha + 1)^2 - 4\beta d_k}}{2\beta},
      \\ 
      \qquad\qquad \text{ for } p_k < 0, \quad l_k > 1, 
    \end{cases}\\
  \end{aligned}
\end{equation}
where
\[
\begin{aligned}
  &
      c_k = (1 - ( p_k\text{ mod }1))( u_k +1) - 1,
    \\
    & S_{l_k} = -\left( u_k - \alpha + 1 \right) p_k + \beta p_k^2,\\
    & l_k = \frac{1 - (S_{l_k}\text{ mod }1)}{ u_k + 1},
    \quad
    d_k = (S_{l_k}\text{ mod 1}) +  u_k.
\end{aligned}
\]
One of the advantages of \eqref{eq:complete-model-ab} 
is that it has the only one steady state at $(u_k=0,p_k=0)$.
For practical purposes, only \emph{locally (asymptotically) stable steady state}, in which the loop returns after small perturbations of its state,
is of interest.

\begin{figure}[ht]
  \centering
  \includegraphics[width=0.85\linewidth]{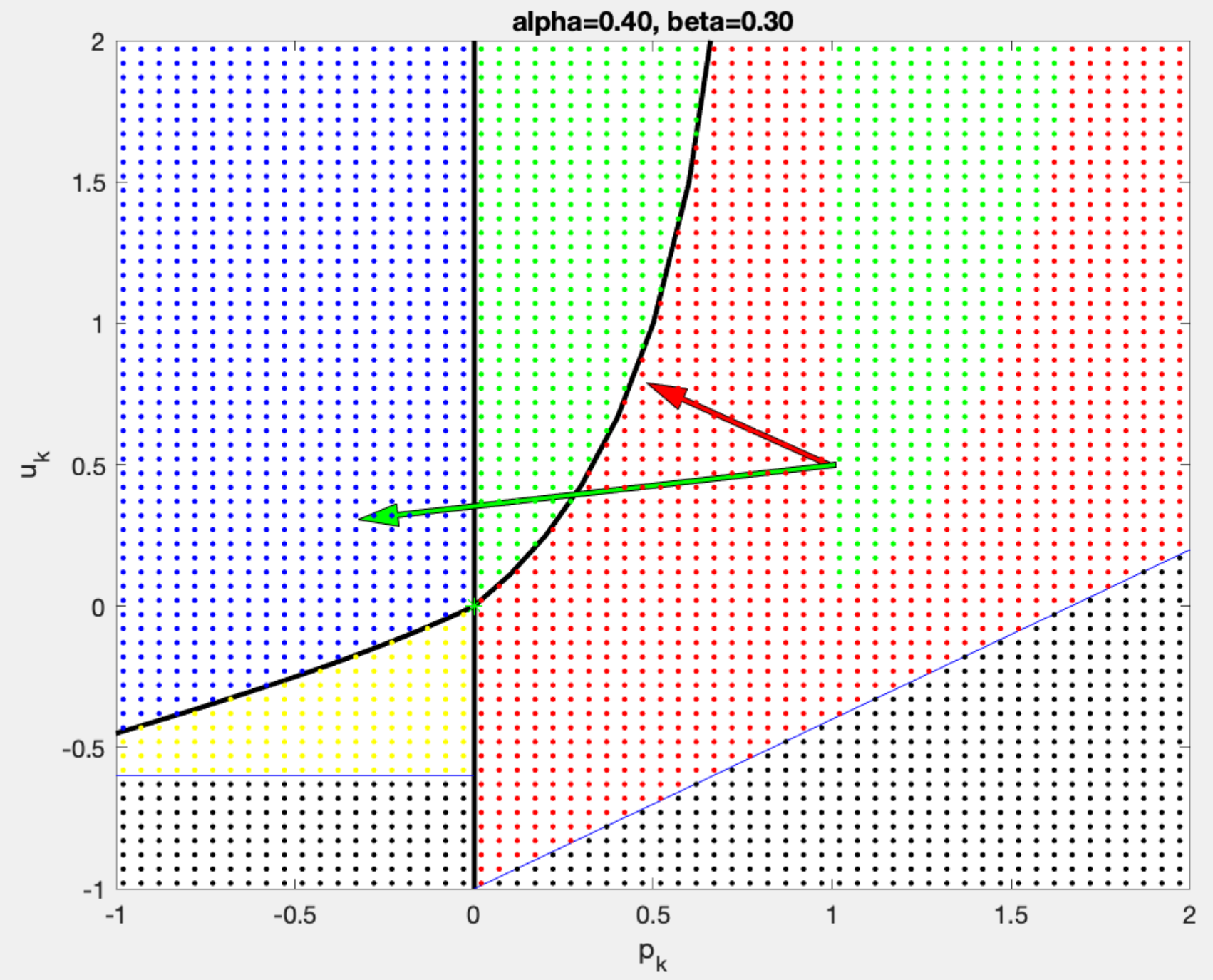}
  \caption{
  Red dots: $p_k>0$, $c_k=c(p_k,u_k)<0$, $p_{k+1}>0$. 
  Green dots: $p_k>0$, $c_k=c(p_k,u_k)>0$, $p_{k+1}<0$.
  Blue dots: $p_k<0$, $l_k=l(p_k,u_k)<1$, $p_{k+1}<0$.
  Yellow dots: $p_k<0$, $l_k=l(p_k,u_k)<1$, $p_{k+1}>0$.
  Black dots: VCO input overload.
  Red arrow: 
  $p_k=0.99,\ u_k=0.5\ \to\ p_{k+1}=0.48$, $u_{k+1}=0.79$.
  Green arrow:
  $p_k=1.01,\ u_k=0.5\ \to\ p_{k+1}=-0.32$, $u_{k+1}=0.31$.
  }
  \label{fig:discont-0}
\end{figure}

\begin{figure*}[ht]
  \centering
  \includegraphics[width=0.48\linewidth]{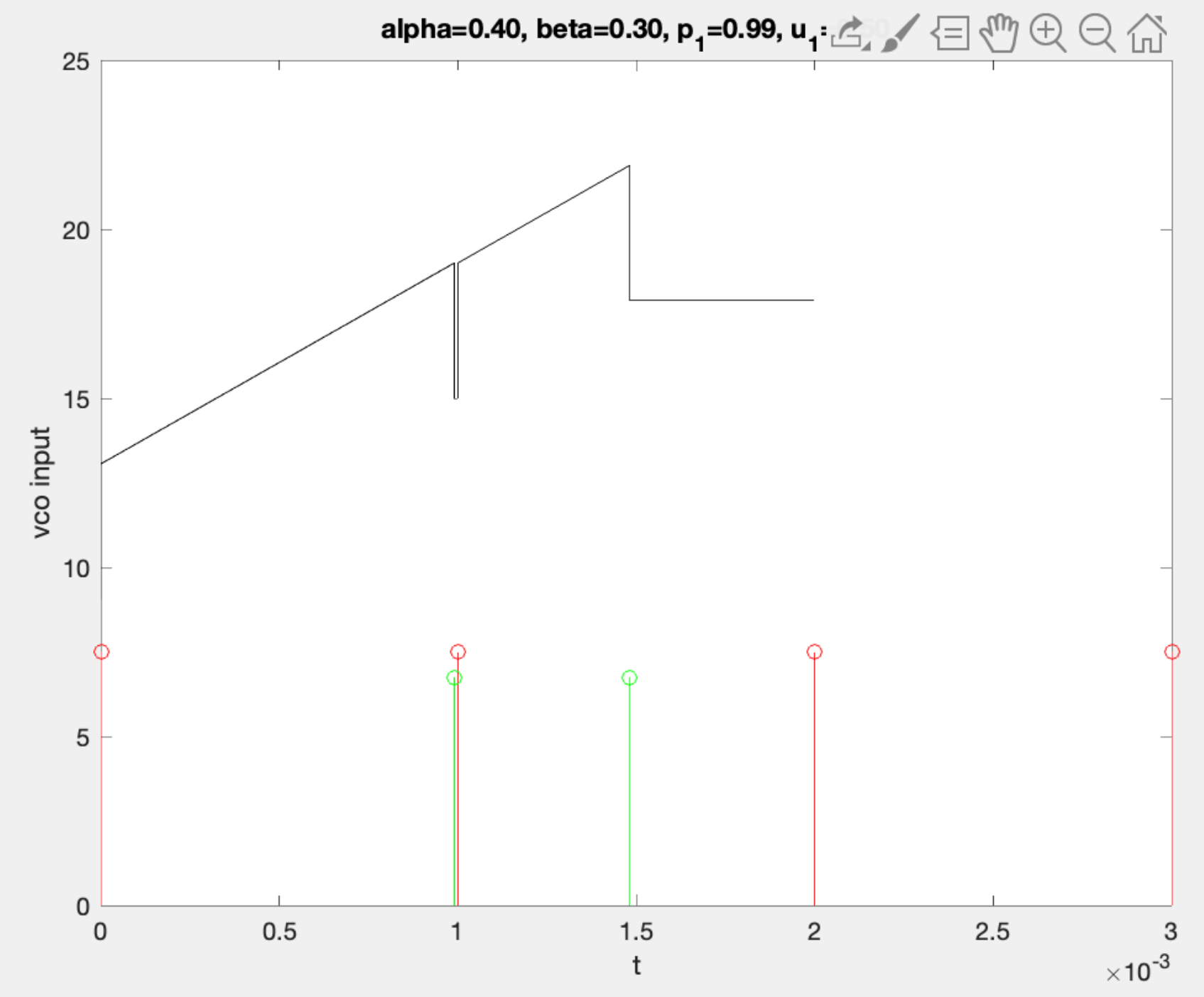}
  \includegraphics[width=0.48\linewidth]{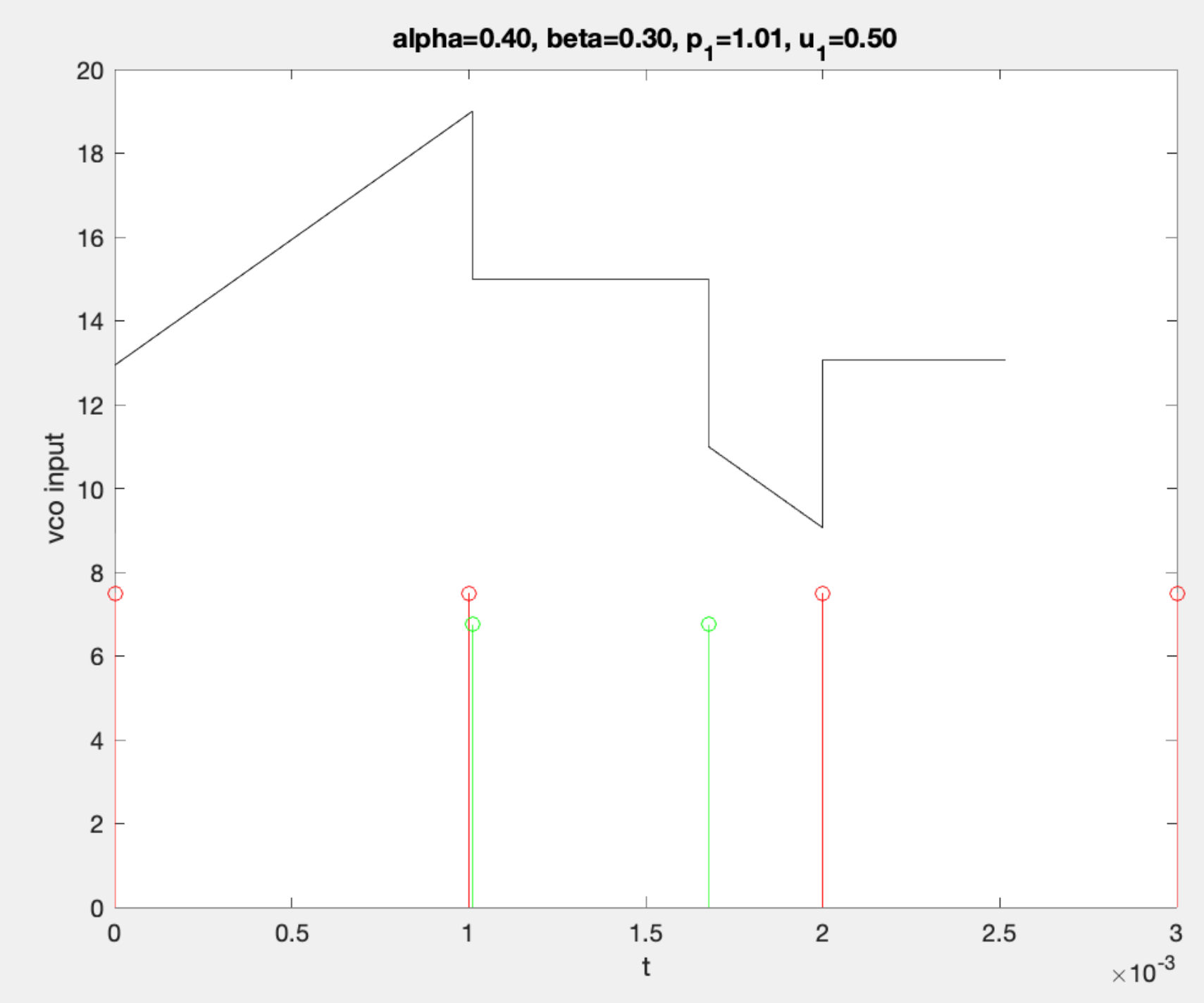}
  \caption{Quantitative difference in PFD behaviour for small difference of phases. 
  Red --- reference trailing edges; green --- VCO trailing edges.
  Black --- VCO input. Left: $p_k=0.99,\ u_k=0.5\ \to\ p_{k+1}=0.48$, $u_{k+1}=0.79$.
  Right:
  $p_k=1.01,\ u_k=0.5\ \to\ p_{k+1}=-0.32$, $u_{k+1}=0.31$}
  \label{fig:discont}
\end{figure*}

Note that the right-hand side of \eqref{eq:complete-model-ab} 
 is continuous and piecewise smooth in  neighborhood of the origin 
and discontinuous far from the origin  
(see Fig.~\ref{fig:discont-0} and the corresponding VCO input signals 
in Fig.~\ref{fig:discont}).
\section{Small-signal analysis: the locked state}
\label{sec:local-stability}
The local stability analysis of the CP-PLL model
via straight-forward linearization may lead to incorrect conclusions because the system is not smooth near the steady state (in fact, it is only piecewise smooth). In \cite{Orla-2007,Orla-2013-review} the stability analysis follows the lines of the Lyapunov approach, however,
the details of the proof are not presented\footnote{``The proof of this assertion is neither trivial nor brief'' \cite[p.9]{Orla-2013-review}.}
and the analysis is done without taking into account the VCO overload. 

Model \eqref{eq:complete-model-ab} has only one steady state 
\begin{equation}
\begin{aligned}
  &  u_k =  u_{k+1} \equiv 0,
  \quad
  p_k =  p_{k+1} \equiv 0,
  \end{aligned}
\end{equation}
which is a locked state if the state is locally asymptotically stable.
In \cite{KuznetsovYYBKKM-2019,KuznetsovYYBKM-2019-arXiv} 
it is shown that a small vicinity of the zero steady state lies outside the VCO overload  if condition $0<\alpha<1$ holds
(thus we can use model \eqref{eq:complete-model-ab} for the local analysis of the loop).

\section{The hold-in range of CP-PLL}
The hold-in range corresponds to the input frequency range,
which allows PLL to keep acquired the locked state
despite small and slow deviations of the input frequency 
or phase.
This notion is similar to the definition
of the hold-in range for classic analog PLLs
\cite{KuznetsovLYY-2015-IFAC-Ranges,LeonovKYY-2015-TCAS,BestKLYY-2016}.
However, as shown below, for the considered CP-PLL model there is always 
a reference signal period $T_{\rm ref}$
such that the steady state $u_k = p_k = 0$ is stable 
(assuming that there is no overload).
Moreover, for any smaller values of $T_{\rm ref}$
the equilibrium $u_k = p_k = 0$ remains locally stable.
Therefore, it is reasonable to give the following definition of the hold-in range for the CP-PLLs.

\begin{definition}
\label{def:hold-in}
The hold-in range of the CP-PLL is a maximum range of 
the input signal periods $T_{\rm ref}=\frac{1}{\omega_{\rm ref}}$:
\begin{equation}
\begin{aligned}
  & 0 < T_{\rm ref} < T_{\text{\rm hold-in}},
\end{aligned}
\end{equation}
such that there exists a locked state 
(i.e., an asymptotically stable steady state of nonlinear discrete time CP-PLL model \eqref{eq:complete-model-ab})
near which the VCO is not overloaded.
\end{definition}

Here, $\omega_{\rm vco}^{\rm free}$ does not affect the hold-in range
and can be predetermined for certainty.
Since it is not possible to choose zero value 
$\omega_{\rm vco}^{\rm free} = 0$ (because in this case 
the transistors inside the charge-pump do not operate properly), 
one can choose  
$\omega_{\rm vco}^{\rm free} = \frac{1}{T_{\text{\rm hold-in}}}$
or equal to the expected frequency of the reference signal.
The stability of the steady state for $0<\alpha<1$ (no local overload condition)
and $0<\beta<2$ 
is proved in Appendix~\ref{app:theorem-proof}.
Then, substituting $T_{\rm ref} = \sqrt{\frac{2C\beta}{K_vI_p}}$ and $T_{\rm ref} = \frac{\alpha}{K_{\rm vco}I_pR}$ from \eqref{eq:alpha-beta-eqs} 
into the above inequalities
we obtain estimate \eqref{eq:hold-in}.
The proof is based on sufficient conditions for stability (as in classical results on stability by the first approximation for nonlinear ODE), at the same time for 
$\beta>2$ the linearized system becomes unstable 
and numerical simulation shows that the origin of nonlinear model \eqref{eq:complete-model-ab} becomes unstable
(see corresponding discussion in Appendix~\ref{app:instability} 
and in Fig.~\ref{fig:unstable-origin}), 
thus the equality is stated in the theorem.

\begin{theorem}
\label{thm.holdin}
The hold-in range of CP-PLL is as follows
  \begin{equation}
  \label{eq:hold-in}
  \begin{aligned}
    \frac{1}{\omega_{\rm ref}} = T_{\rm ref} < T_\text{\rm hold-in} 
    =
    \min\bigg\{
    \sqrt{\frac{4C}{K_{\rm vco}I_p}},
    \frac{1}{K_{\rm vco}I_pR}
    \bigg\}.
  \end{aligned}
\end{equation}
\end{theorem}

\begin{figure}[!t]
\centering
\subfloat[origin is unstable for $\beta>2$. Simulation of model \eqref{eq:complete-model-ab}, $\alpha=0.5, \beta=0.1$]{
  \includegraphics[width=0.42\linewidth]{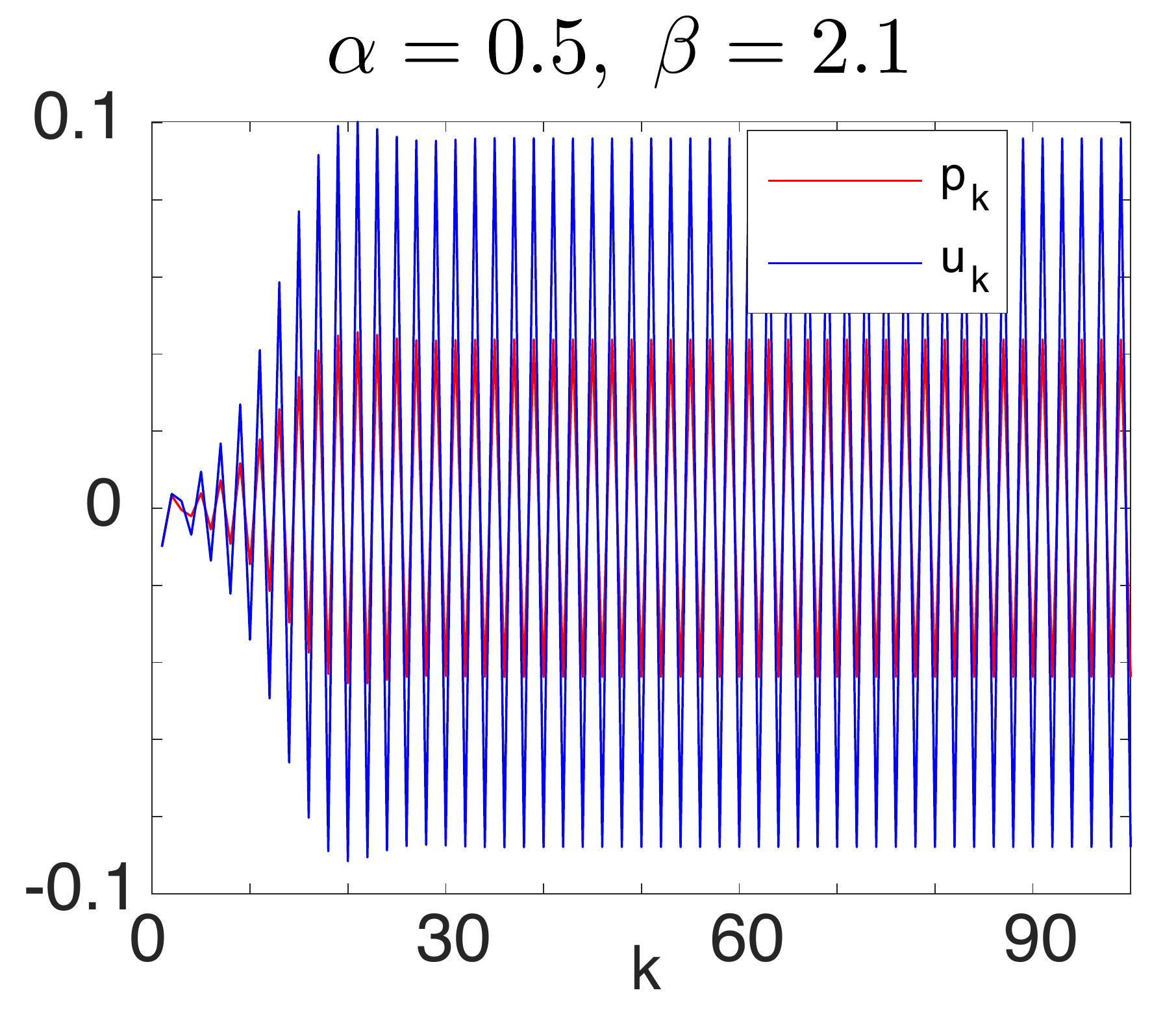}
}
\quad
\subfloat[origin is unstable for $\alpha>1$ simulation of model taking into account overload from \cite{KuznetsovYYBKM-2019-arXiv}, $\alpha=1.1, \beta=1.9$]{
  \includegraphics[width=0.42\linewidth]{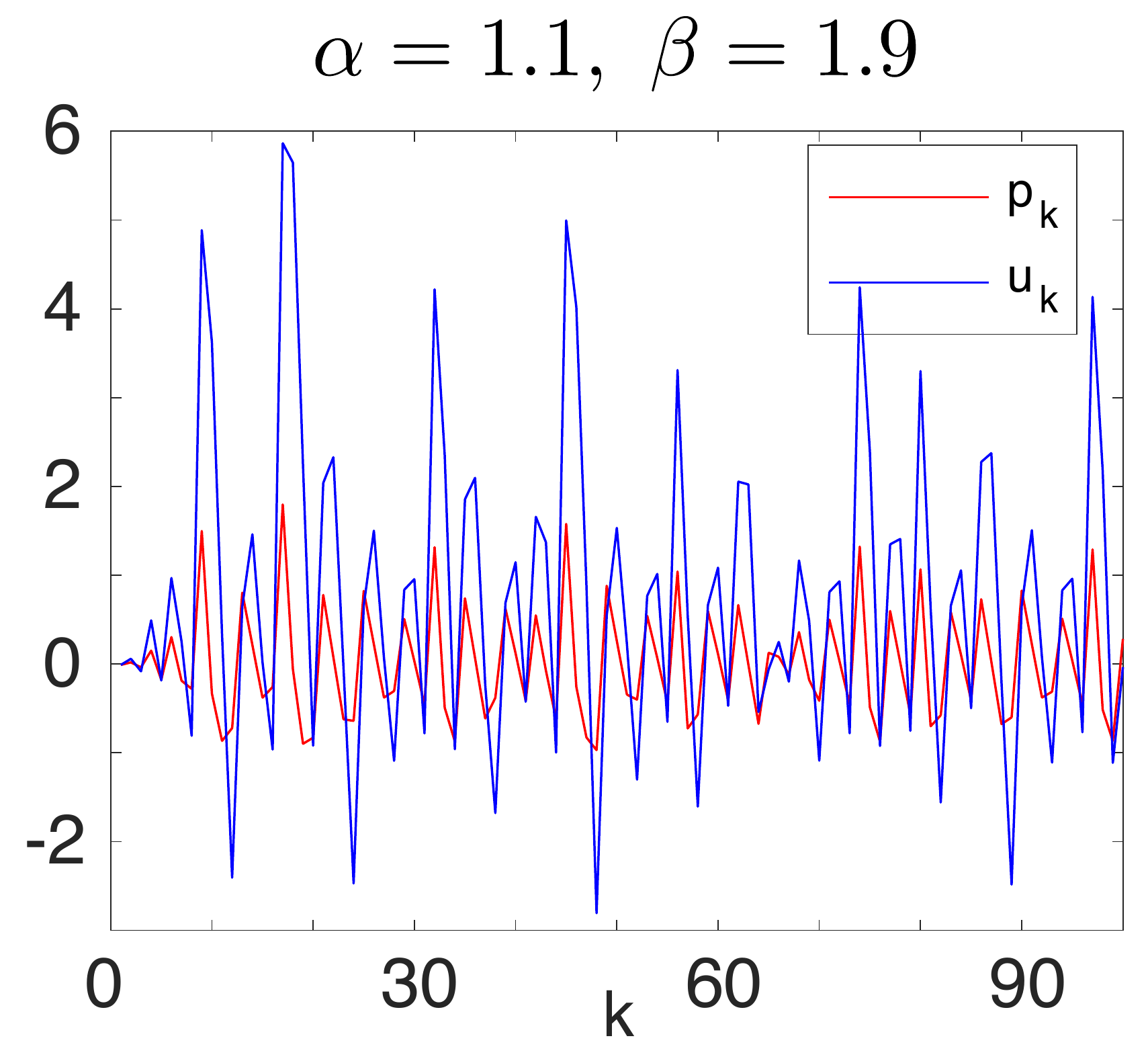}
}
\caption{Numerical simulation of local stability, $p_0=u_0=-0.01$.}
\label{fig:unstable-origin}
\end{figure}

Remark that \eqref{eq:hold-in} refines the estimate 
that can be obtained according to Definition 1 
from the results in \cite{Orla-2013-review}, 
which do not take into account the overload: $T_\text{\rm hold-in} \geq \sqrt{\frac{4C}{K_{\rm vco}I_p}}$.

\section{The pull-in range and non-local analysis}
\label{sec:pull-in}

Following the Gardner's conjecture on CP-PLL \cite[p.1856]{Gardner-1980} 
and the Egan conjecture on the pull-in range of type 2 APLL 
\cite[p.176]{Egan-1981-book},\cite{KuznetsovLYYK-2020-IFACWC,KuznetsovLYY-2021-TCASII}, in \cite[p.6]{Fahim-2005} it was noted that in order to have an infinite pull-in range in CP-PLL, an active filter must be used for the loop filter (i.e. in this case the pull-in range is only limited by 
the VCO tuning range \cite[p.32]{Shu-2005},\cite[p.29]{Razavi-1996}).
However, unlike classic PLLs with PI filter \cite{AlexandrovKLNS-2015-IFAC-Pull-in,KuznetsovLYY-2019-DAN},
for some parameters and initial input frequencies the CP-PLL cannot acquire the locked state due to the presense of nontrivial oscillations (attractors) 
in the phase space. 

\begin{figure}[ht]
  \centering
  \includegraphics[width=0.8\linewidth]{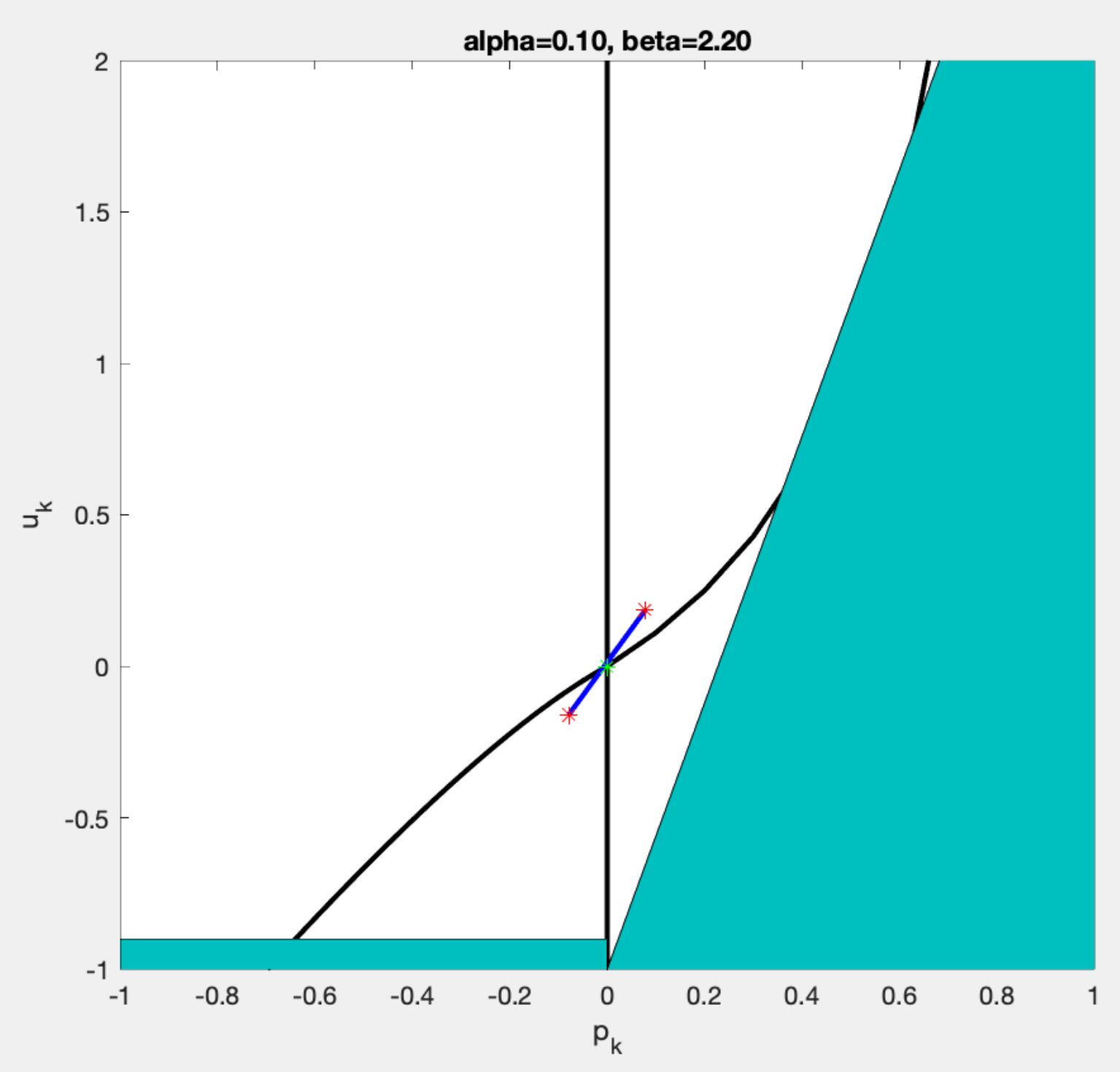}
  \caption{Period-2 limit cycle. Red stars: limit cycle; turquoise: overload; black curves: separation of the smoothness regions.}
  \label{fig:2-cycle}
\end{figure}

\begin{figure}[ht]
  \centering
  \includegraphics[width=0.6\linewidth]{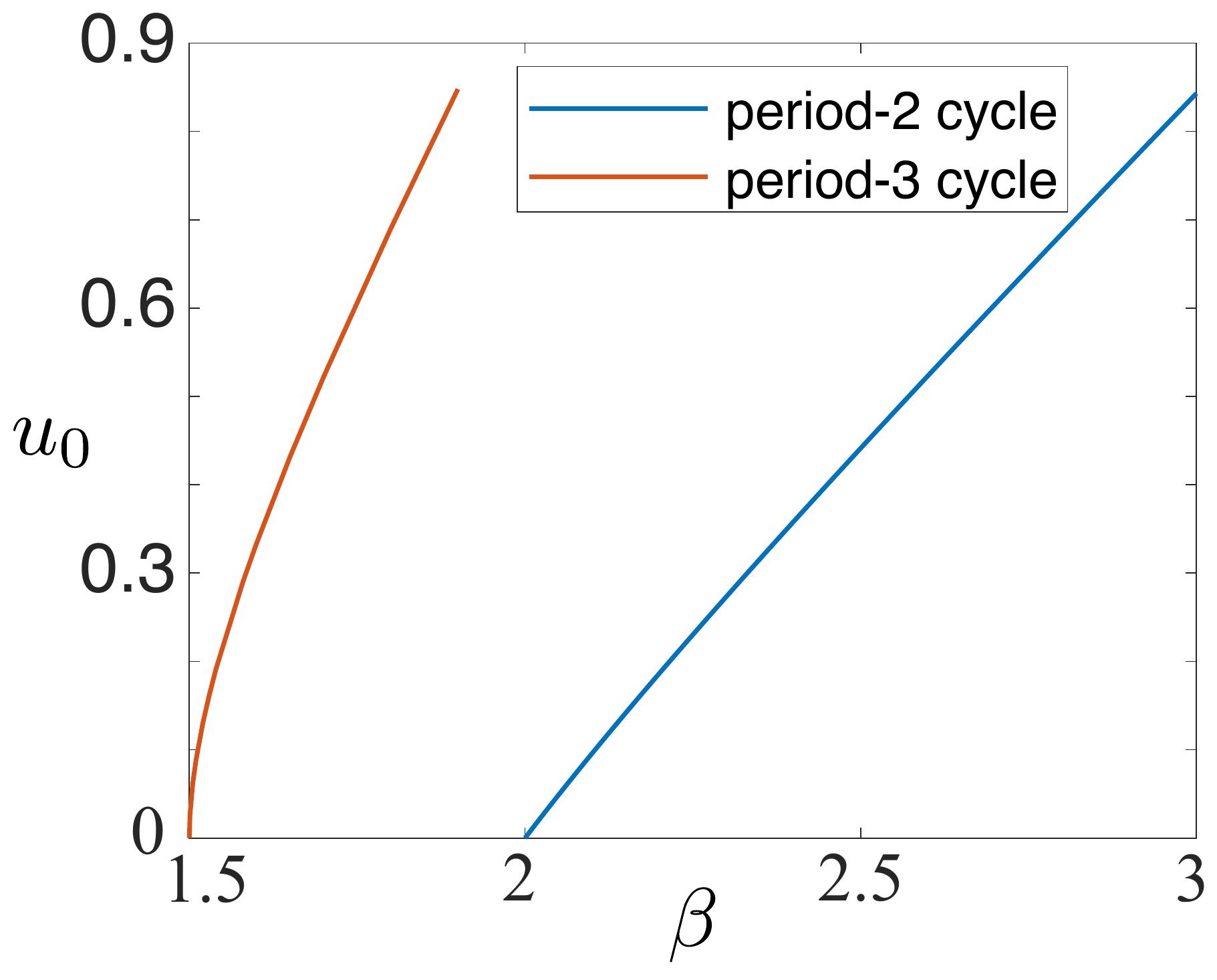}
  \caption{
  Initial value $u_0$ corresponding to period-2 and period-3 cycles.
  }
  \label{fig:period-2-3-bif}
\end{figure}

In discrete time model \eqref{eq:complete-model-ab} the period-2 limit cycles have the form:
\begin{equation}
  \begin{aligned}
    & p_0=\frac{- \sqrt\beta+\sqrt{9\beta - 16} }{4\sqrt\beta}>0,
    \quad
    u_0 = \frac{2p_0}{1-2p_0},\\
    & p_1 = -p_0,
    \quad
    u_1 = u_0 +2\beta p_1.
  \end{aligned}
\end{equation}
 These period-2 limit cycles do not exist for $0<\beta<2$ (see Fig.~\ref{fig:period-2-3-bif}, Fig.~\ref{fig:2-cycle}).

\begin{figure}[ht]
  \centering
  \includegraphics[width=0.7\linewidth]{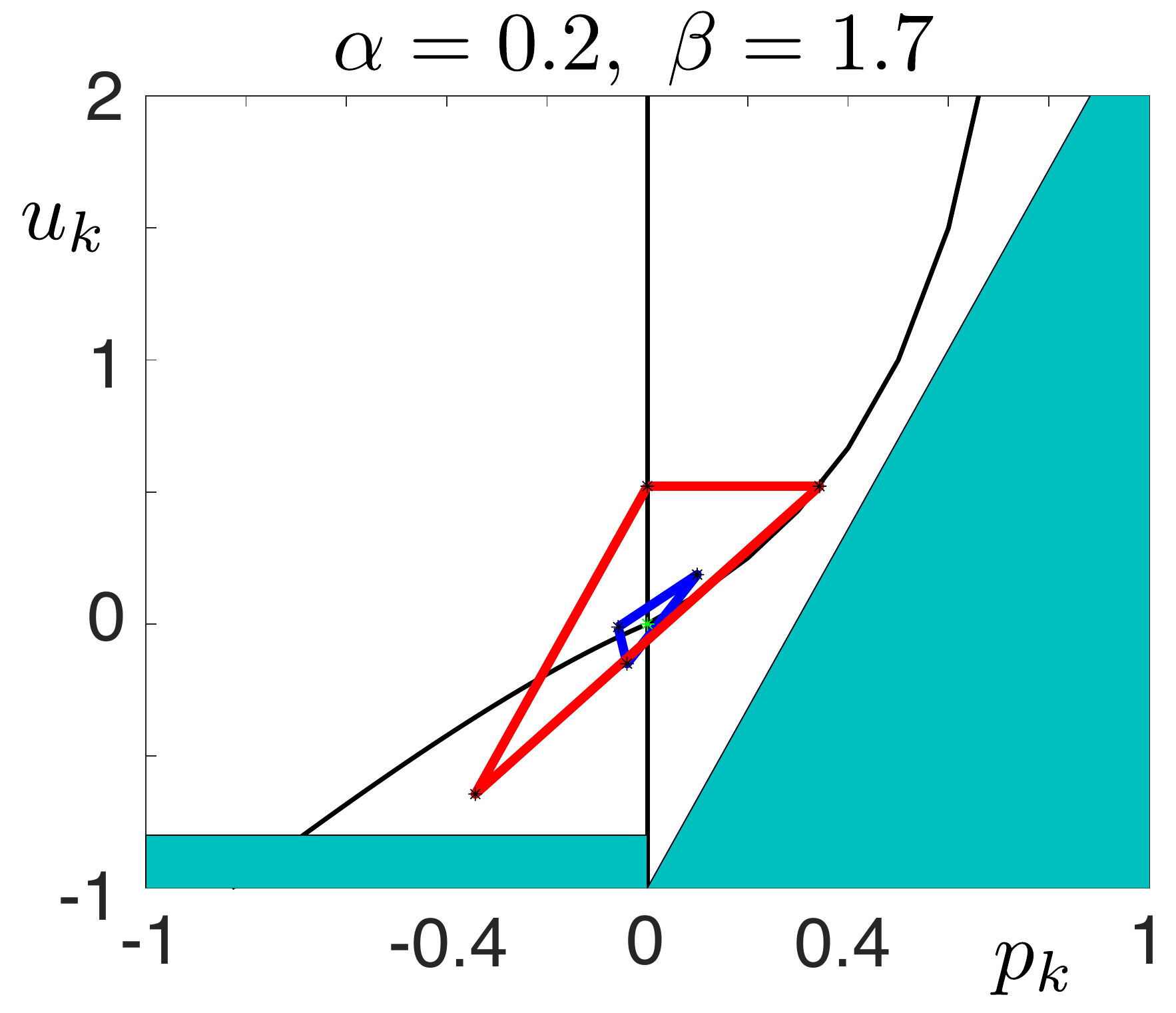}
  \caption{Stable (red) and unstable (blue) period-3 limit cycles (hidden oscillations). Turquoise: overload; black curves: separation of the smoothness regions.}
  \label{fig:3-cycle}
\end{figure}

Also, there are period-3 limit cycles of the form (see Fig.~\ref{fig:3-cycle})
\begin{equation}
  \label{eq:period-3-limit-cycle}
  \begin{aligned}
    & p_0 = 0, 
    \quad u_0 =\frac{2\beta-3+\sqrt{2\beta}\sqrt{2\beta-3}}{3},\\
    & p_1 = -\frac{u_0}{u_0+1},
    \quad u_1 = u_0-2\beta\frac{u_0}{1+u_0},\\
    & p_2 =-p_1= \frac{u_0}{1+u_0},
    \quad u_2 = u_0,
  \end{aligned}
\end{equation}
where it is assumed that $u_1>1-\alpha$ and $u_2>2\beta p_2-1$ to avoid the VCO overload.
Note that these cycles exist only for $\beta>\frac{3}{2}$, 
and for $\beta=\frac{3}{2}$ they merge with the origin.
Also unstable limit cycles may exist in the model (see, e.g. Fig.~\ref{fig:3-cycle}).

While for the discrete time model \eqref{eq:complete-model-ab} 
the limit cycles of low-periods without overload 
can be easily found analytically (see, e.g. \cite{Razavi-hb-2016}), 
the computing of limit cycles in the case of higher periods or overload 
leads to complicated equations which need to be solved numerically.
Since the steady state is stable for $0<\beta<2$, $0<\alpha<1$,
the existing limit cycles for such parameters can be classified 
as hidden oscillations \cite{LeonovK-2013-IJBC,Kuznetsov-2020-TiSU}.

\subsection{The pull-in range estimate}
For given parameters the input frequency range, for which a locked state is acquired from any possible initial state, is known as the pull-in range.

\begin{definition}
The pull-in range of CP-PLL is a maximum range of 
the input signal periods $T_{\rm ref}=\frac{1}{\omega_{\rm ref}}$
within the hold-in range:
\begin{equation}
  0 <  T_{\rm ref}
      <  T_{\text{\rm pull-in}}
      \leq  T_{\text{\rm hold-in}},
\end{equation}
such that for any initial state the CP-PLL 
acquires a locked state.
\end{definition}

\begin{figure}[ht]
  \centering
  \includegraphics[width=0.9\linewidth]{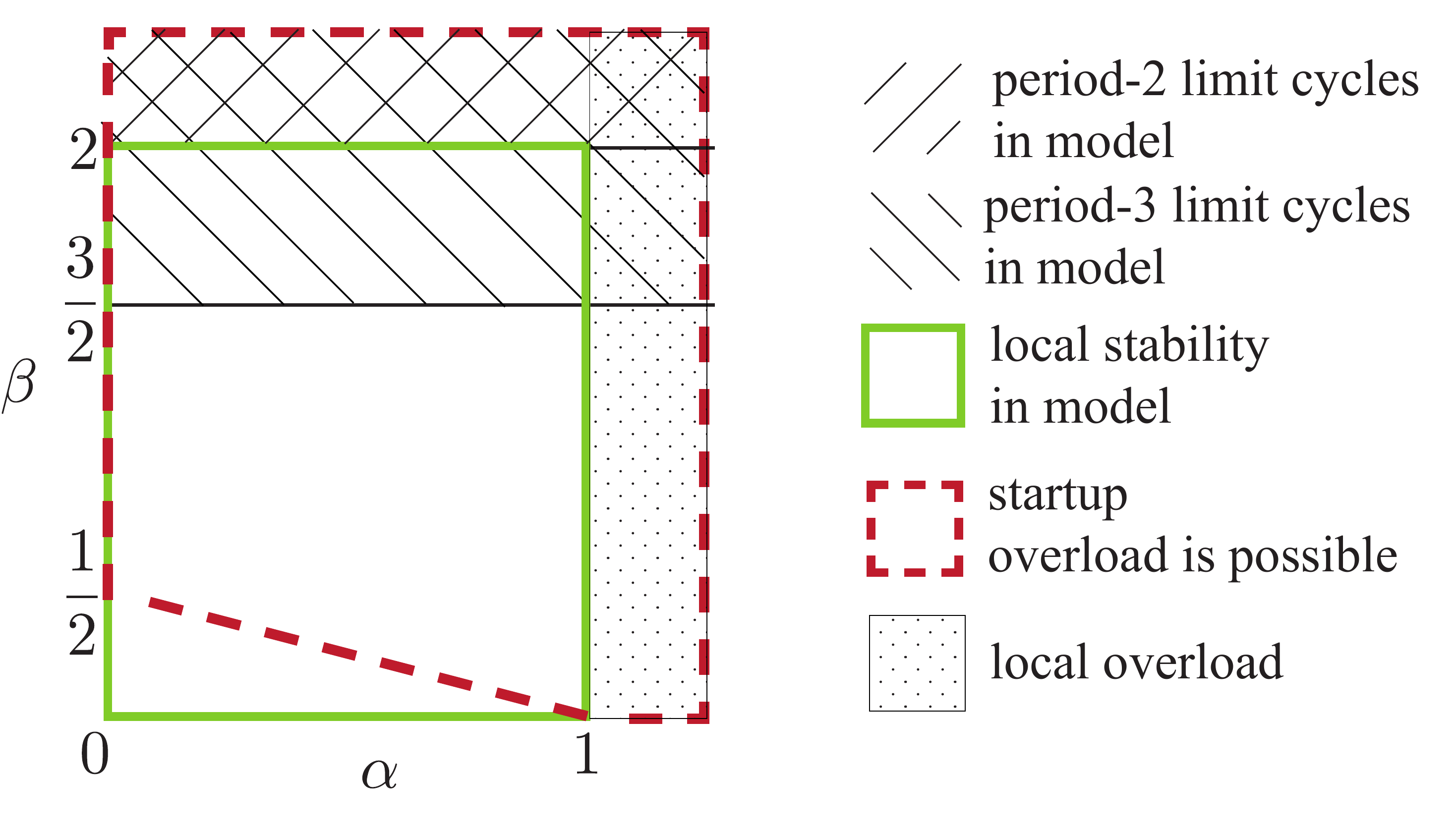}
  \caption{Stability regions in parameter space}
  \label{fig:stability-rectangle}
\end{figure}

Since period-2 and period-3 limit cycles exist for $\beta>\frac{3}{2}$ and 
$\alpha <1$, using $T_{\rm ref} = \sqrt{\frac{2C\beta}{K_vI_p}}$ and $T_{\rm ref} = \frac{\alpha}{K_{\rm vco}I_pR}$ from \eqref{eq:alpha-beta-eqs},
the following upper estimate of the pull-in range can be obtained 
(see Fig.~\ref{fig:stability-rectangle}):

\begin{theorem}\label{thm.mainpull}
\begin{equation}
  \label{eq:pull-in-estimate}
  \begin{aligned}
    & T_{\text{pull-in}}\leq
    \min\bigg\{
    \sqrt{\frac{3C}{K_{\rm vco}I_p}},
    \frac{1}{K_{\rm vco}I_pR}
    \bigg\}.
  \end{aligned}
\end{equation}
\end{theorem}

The existence of hidden periodic oscillations with higher periods may 
further restrict the pull-in range\footnote{
For the classical analog PLL the birth of hidden oscillation
(without loss of stability for the locked states)
may cause the loss of global stability (hidden boundary of global stability) and restrict the pull-in range 
\cite{BianchiKLYY-2016,KuznetsovLYY-2017-CNSNS,Kuznetsov-2020-TiSU,KuznetsovLYYKKRA-2020-ECC}).}.

\begin{figure*}[ht]
  \centering
  \includegraphics[width=\linewidth]{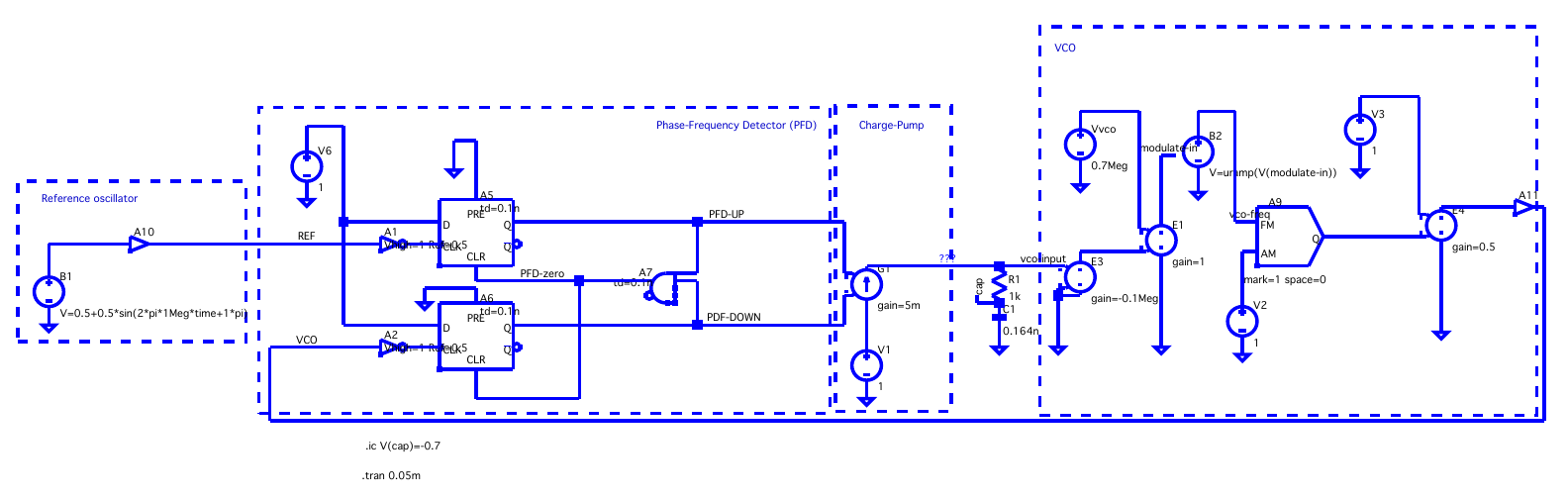}
  \caption{LTspice schematic of CP-PLL. Circuit parameters:$R=1$k$\Omega$,
$C=0.114$~nF,
$\omega_{\rm ref}=1$~MHz,
$K_{\rm vco} = 0.1$~MHz/V,
$\omega_{\rm vco}^{\rm free} = 0.7$~MHz,
$v_c(0)=10$ V,
$\tau_0=0.345\cdot 10^-6$ s.}
  \label{fig:spice-schematic}
\end{figure*}

\begin{figure}[!t]
\centering
\subfloat[$\beta = 1.4$, $C=0.179$ nF]{
  \includegraphics[width=0.49\linewidth]{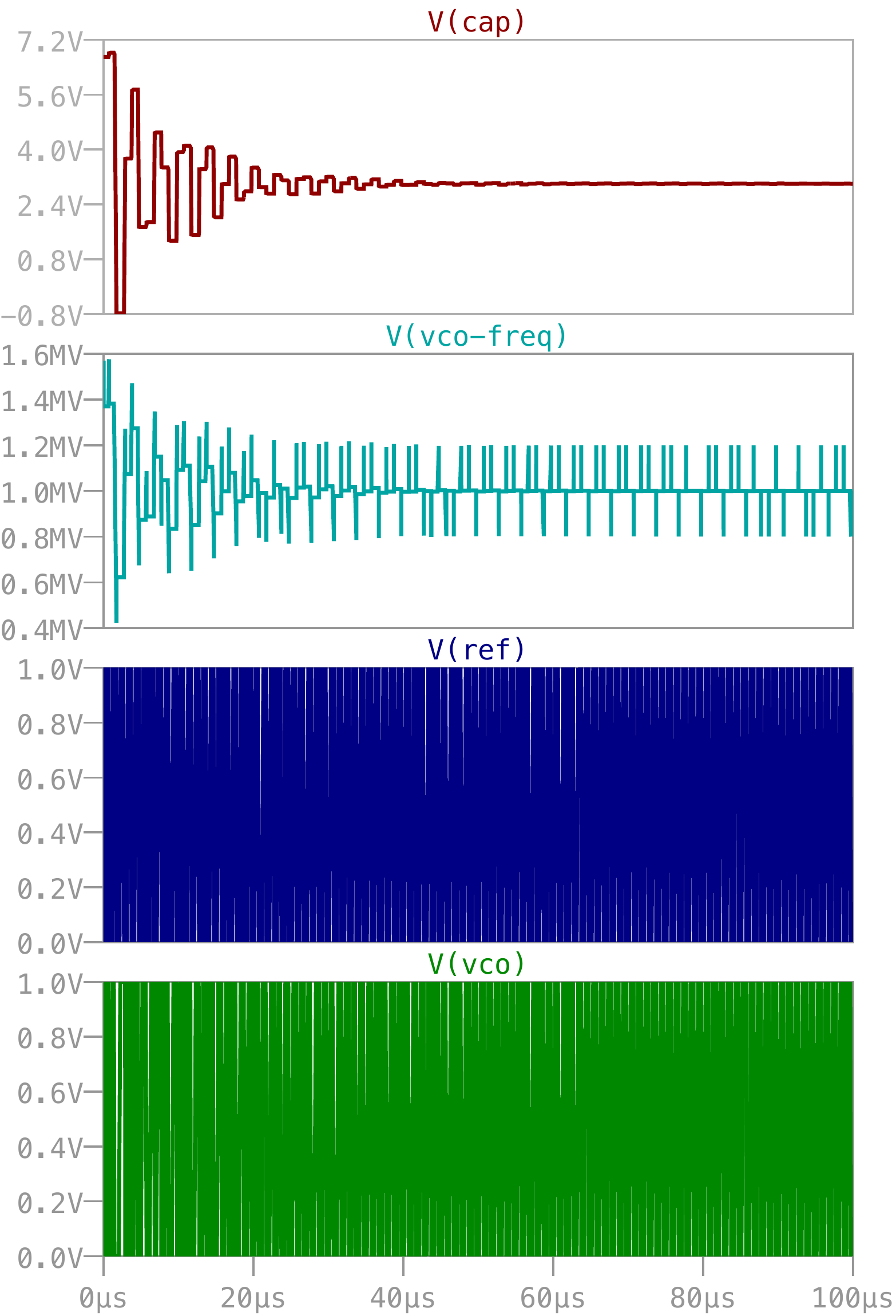}
}
\subfloat[$\beta = 1.6$, $C=0.156$ nF]{
  \includegraphics[width=0.49\linewidth]{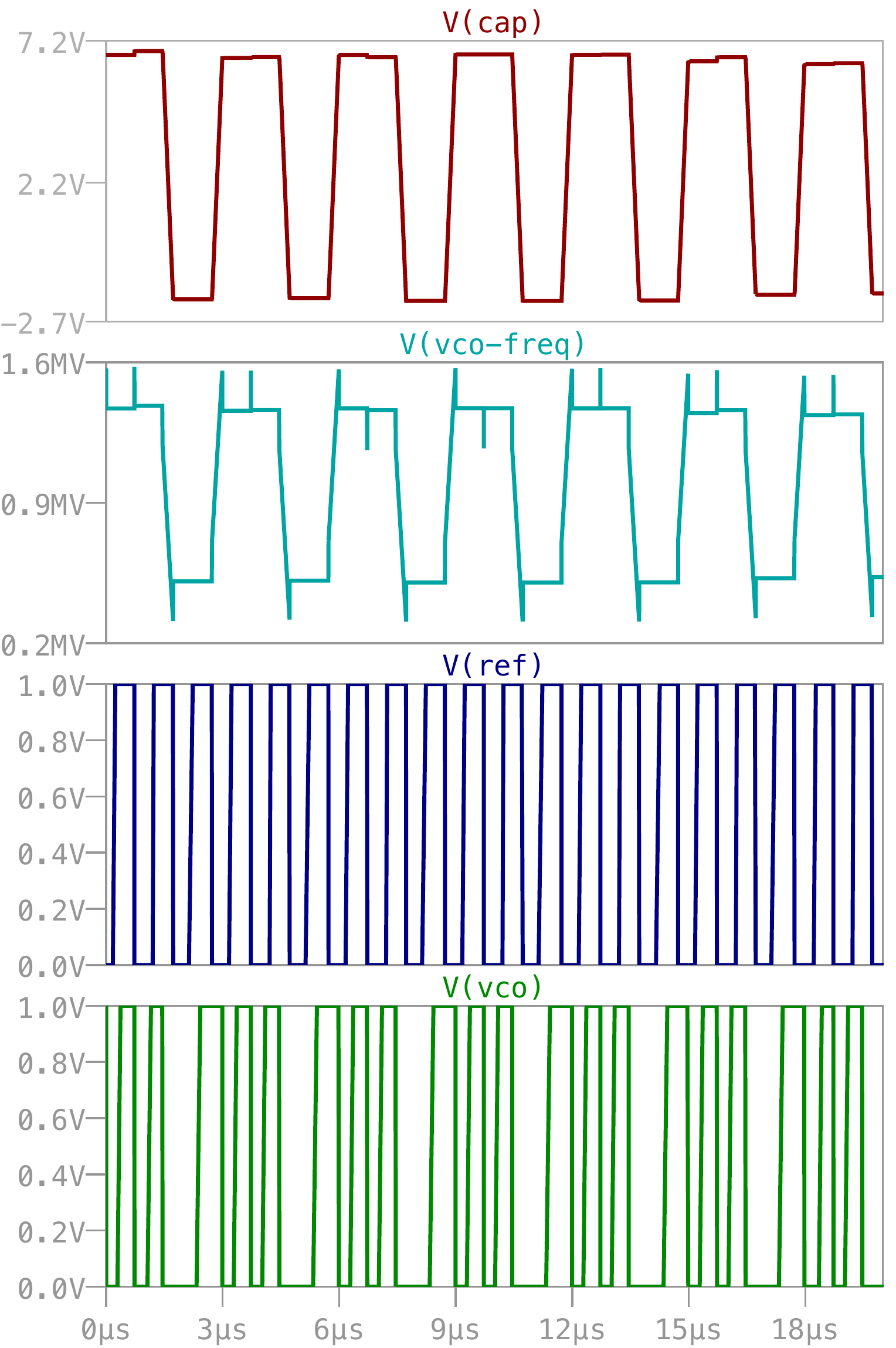}
}
\caption{Birth of a stable period-3 limit cycle in 
LTspice for $\beta=1.5$. Reference: ``V=0.5+0.5*sin(2*pi*1Meg*time-0.45*pi)'', $\omega_{\rm ref}=1$~MHz,
$K_{\rm vco} = 0.1$~MHz/V, $\omega_{\rm vco}^{\rm free}=0.7$~MHz, 
$v_c(0)=6.7$~V, $R=0.4$~k$\Omega$, $\alpha = 0.2$.}
\label{fig:spice-sim-cycle}
\end{figure}

\begin{figure}[!t]
\centering
\subfloat[$\alpha = 0.9$, $R=1.8$ k$\Omega$.
]{
  \includegraphics[width=0.49\linewidth]{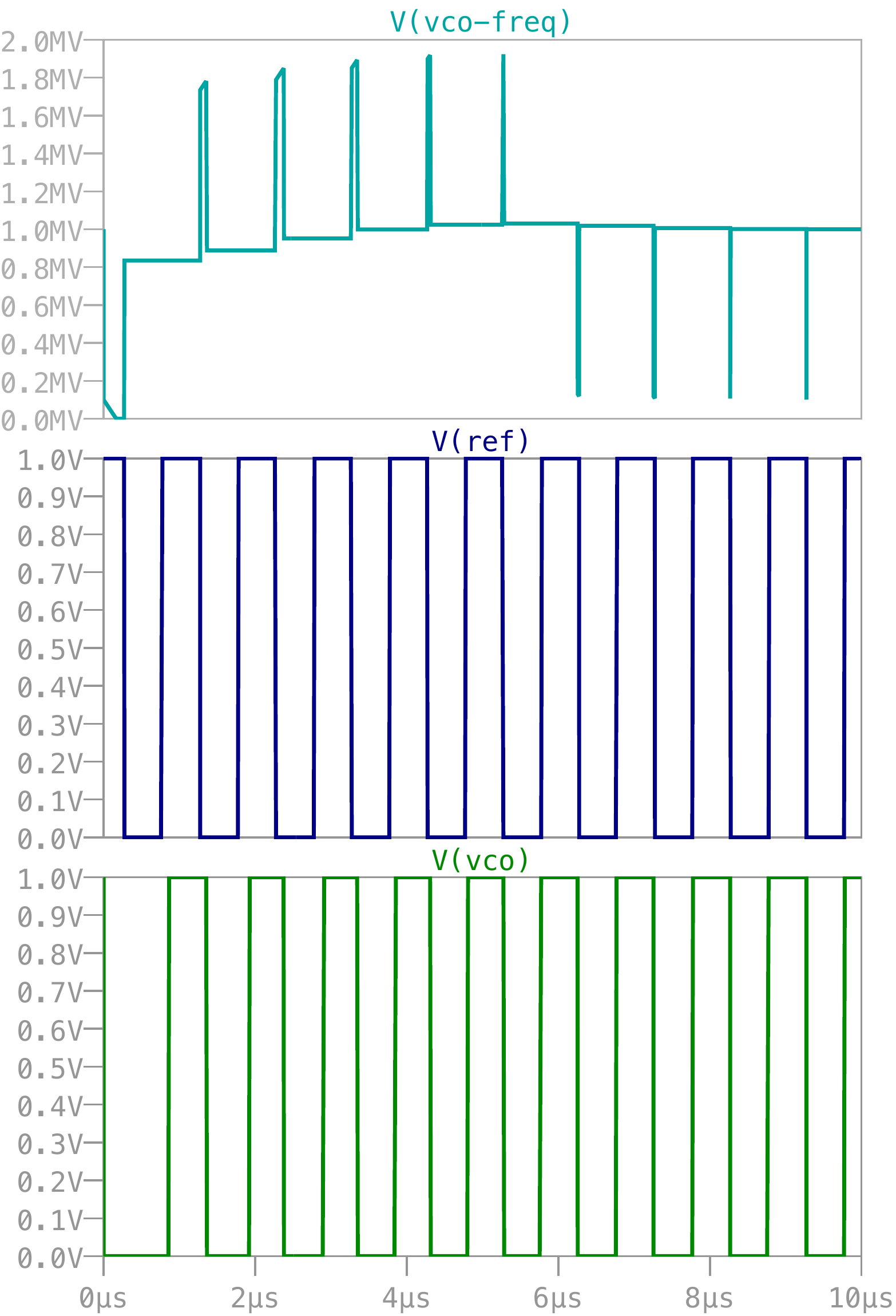}
}
\subfloat[$\alpha = 1.1$, $R=2.2$ k$\Omega$.
]{
  \includegraphics[width=0.49\linewidth]{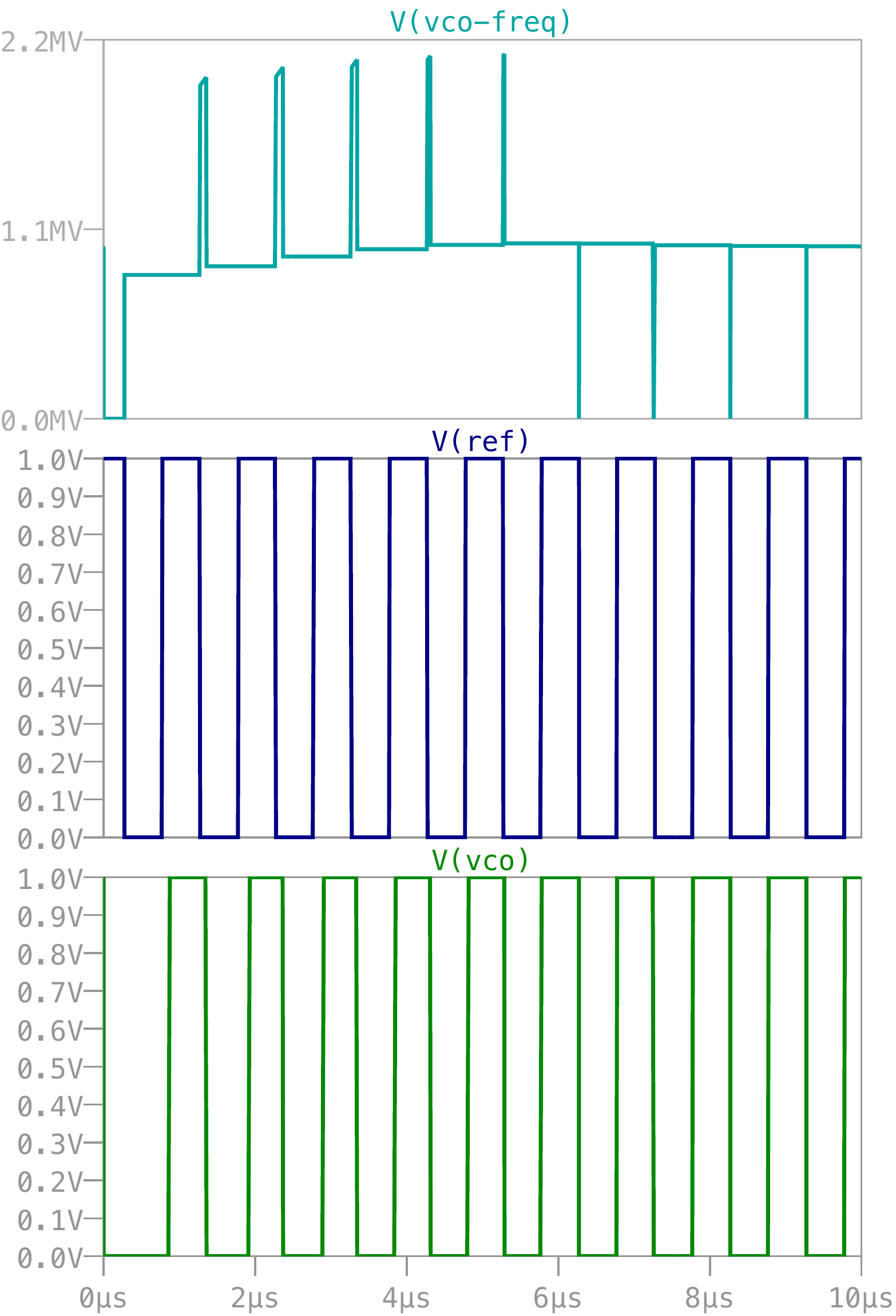}
}
\caption{Local overload in LTspice for $\alpha>1$: bursts in subfigure (b) are too large which leads to VCO overload in synchronized state. Reference: ``V=0.5+0.5*sin(2*pi*1Meg*time+0.45*pi)'',$\omega_{\rm ref}=1$~MHz,
$K_{\rm vco} = 0.1$~MHz/V, $\omega_{\rm vco}^{\rm free}=1$~MHz, $v_c(0)=6.7$ V, $C=0.156$ nF, $\beta = 0.3$.}
\label{fig:spice-sim-local-overload}
\end{figure}

\begin{figure}[!t]
\centering
\subfloat[$v_c(0)=-4.5$ V.
]{
  \includegraphics[width=0.45\linewidth]{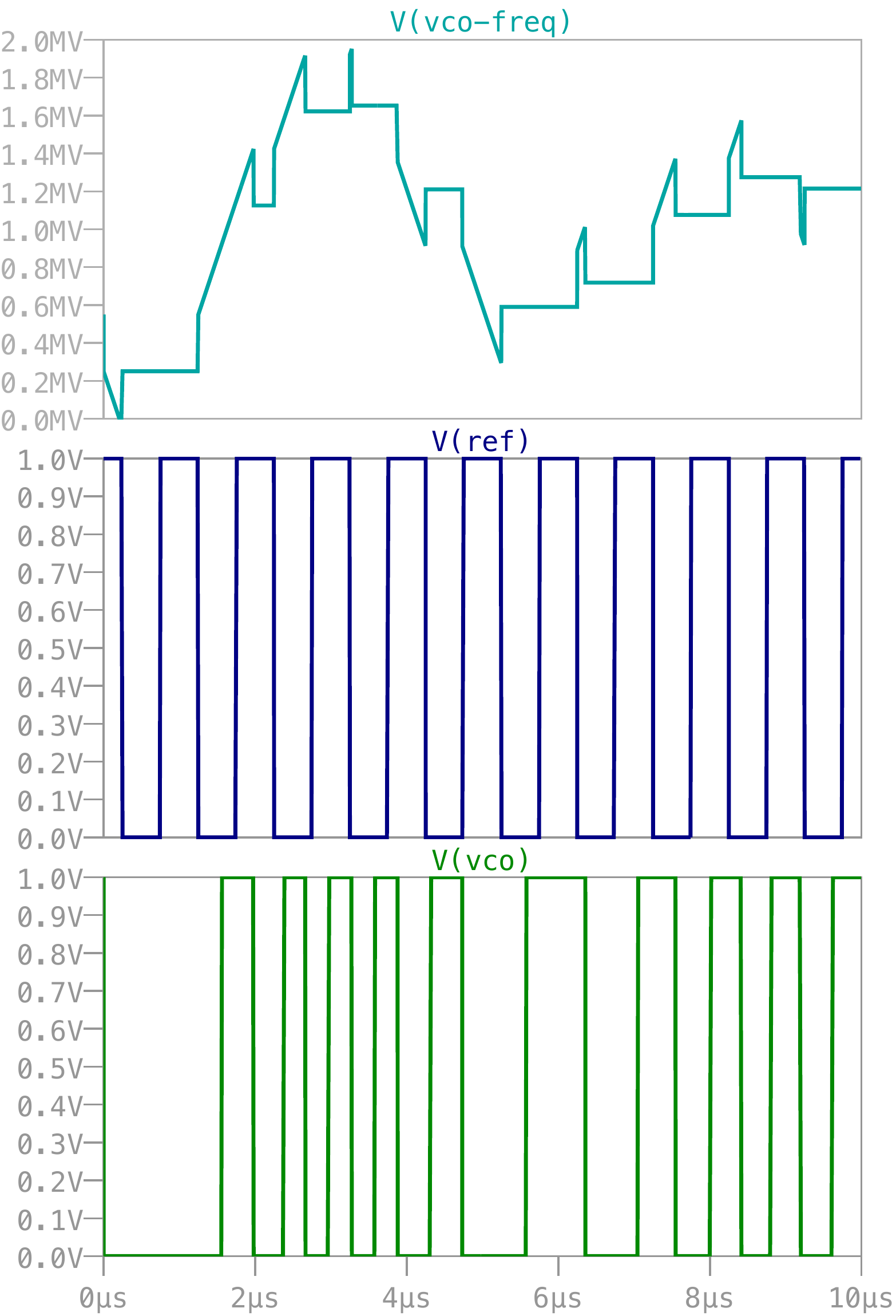}
}\quad
\subfloat[ $v_c(0)=-11$ V.
]{
  \includegraphics[width=0.45\linewidth]{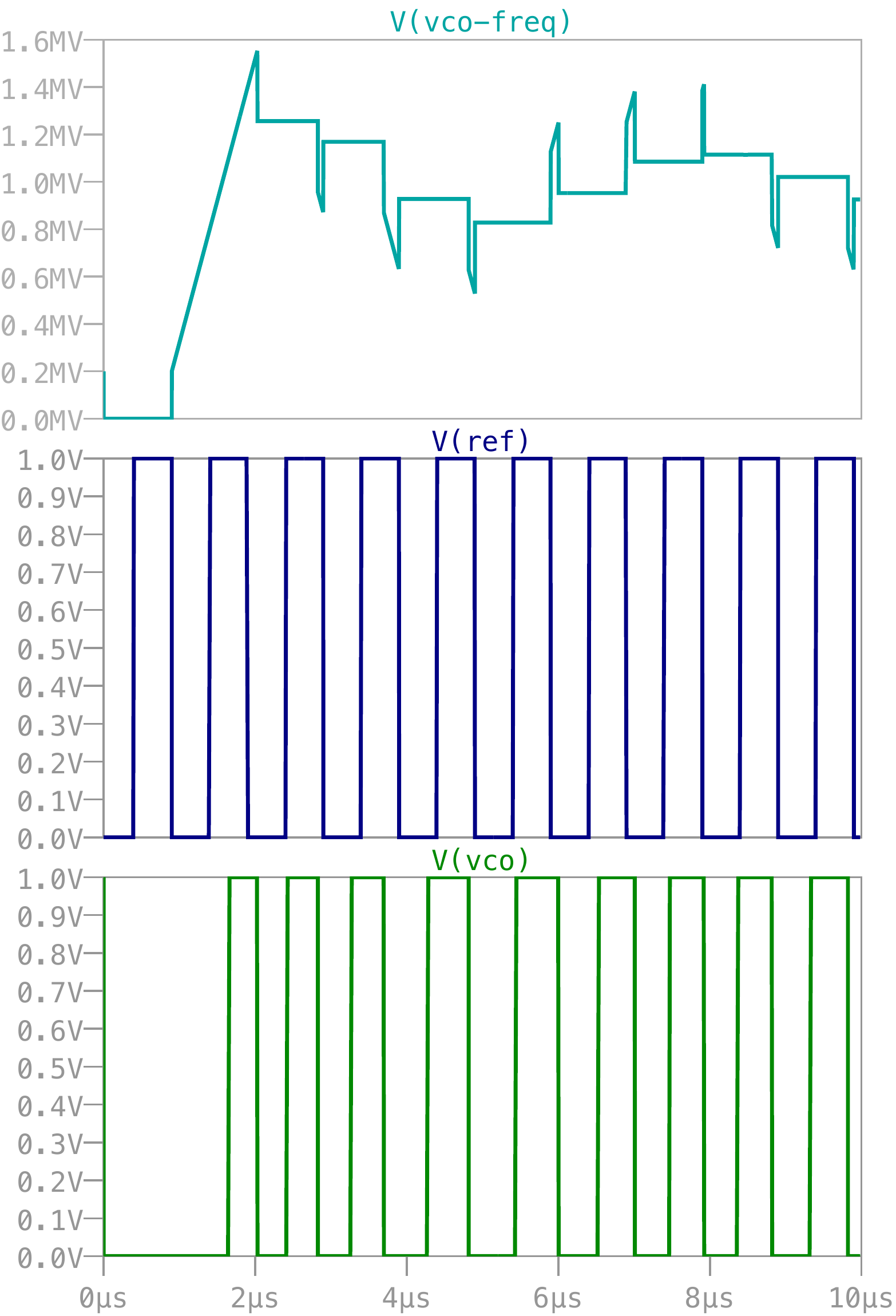}
}
\caption{Non-local overload at startup in LTspice. $\omega_{\rm ref}=1$~MHz, 
$\omega_{\rm ref}=1$~MHz,
$K_{\rm vco} = 0.1$~MHz/V, $\omega_{\rm vco}^{\rm free}=1$~MHz,
$\alpha = 0.3$, $\beta=0.6$, $R=0.6$ k$\Omega$, $C=0.417$~nF.
}
\label{fig:spice-sim-non-local-overload}
\end{figure}

Stability analysis is confirmed by the simulation in LTspice (see Fig.~\ref{fig:spice-schematic}).
In Fig.~\ref{fig:spice-sim-cycle}, we show that period-3 limit cycle emerges for $\beta>1.5$.
In Fig.~\ref{fig:spice-sim-local-overload}, we show that VCO overload may happen in the synchronized state for $\alpha > 1$. For $\alpha<1$ there is no overload.
In Fig.~\ref{fig:spice-sim-non-local-overload} we show possibility of non-local overload at startup.

\section{Conclusions}
The stability of the Charge-Pump PLL has been studied for a long time.
However, in these studies the VCO input overload was not fully taken into account and such important engineering parameters as the hold-in and pull-in ranges were not introduced and estimated.
In this work, the definitions of the hold-in and pull-in ranges  for the CP-PLL are introduced in terms of the input signal period and frequency and the corresponding estimations are discussed.
We show that the VCO input overload limits the hold-in and pull-in ranges even more than the domains of parameters corresponding to linear stability and non-existence of limit cycles.
The study provided an answer to Gardner’s conjecture on the similarity of transient responses of CP-PLL and equivalent classical PLL 
\cite[p.1856]{Gardner-1980} and to conjectures on the pull-in range of CP-PLL 
with proportionally-integrating filter \cite[p.6]{Fahim-2005},\cite[p.32]{Shu-2005},\cite[p.29]{Razavi-1996}.

\section*{Acknowledgment}
The work is supported by the Russian Science Foundation (project 19-41-02002), and in part by the Leading Scientific Schools of Russia Project
under Grant NSh-2624.2020.1 (open access).
Authors would also like to thank Giovanni Bianchi (Advantest corporation) for valuable comments and discussions on practical realizations  of CP-PLL.

\begin{appendices} 
\section{Local stability for piecewise-smooth systems}
\label{app:stability-theory}
This section is organized in the following way.
Subsection~\ref{app:piecewise-smooth systems} considers definitions of the piecewise smooth functions.
Subsection~\ref{app:linearization} states that piecewise-linear functions approximate well piecewise-smooth functions in the neighborhood of the steady state.
Finally, subsection~\ref{app:stability} proves that stability of the piecewise linearized system implies stability of the steady state for the original piecewise smooth system.

\subsection{Piecewise-smooth systems}
\label{app:piecewise-smooth systems}
\begin{definition}
\label{def:non-tangential}
A {\em hypersurface} $S \subset \br^n$ is a $(n-1)$-dimensional smooth manifold embedded in $\br^n$. Hypersurfaces $S_1, \ldots, S_k$ are said to be {\em mutually non-tangential} on a subset $O \subset \br^n$ if the following claim holds:
\begin{enumerate}
\item[i)] Whenever $x_\ast \in S_{i_1} \cap \ldots \cap S_{i_s} \cap O$ for some subset $\{i_1, \ldots, i_s\} \subset \{1, \ldots,k\}$ of this set ($i_p \neq i_q \; \forall p \neq q$), the normals to $S_{i_1}, \ldots, S_{i_s}$ at the point $x_\ast$ are linearly independent.
\end{enumerate}
\end{definition}

\begin{definition}
\label{def:piecewise-smooth-function}
A function $f: O \subset \br^n \to \br^n$ is said to be {\em piecewise smooth in a neighborhood of a point} $x_\ast \in O$ if there exists an open vicinity $\mathcal{O} \subset O$ of $x_\ast$ and a set $S_1, \ldots, S_k$ of hypersurfaces such that the following claims are true:
\begin{enumerate}
\item[i)] The function $f(\cdot)$ is continuous on $\mathcal{O}$;
\item[ii)] The function $f(\cdot)$ is continuously differentiable on any connected component $\mathfrak{C}$ of the set $\mathcal{O} \setminus \big[ S_1 \cup \ldots \cup S_k\big]$ and moreover, can be extended from $\mathfrak{C}$ to an open neighborhood of the closure $\overline{\mathfrak{C}}$ with retaining this smoothness;
\item[iii)] The hypersurfaces $S_1, \ldots, S_k$ are mutually non-tangential on $\mathcal{O}$.
\end{enumerate}
\end{definition}
The following is the main result of the section.

\subsection{Linearization of the piecewise-smooth function}
\label{app:linearization}
\begin{definition}
\label{def.1}
A function $q: \br^n \to \br^n$ is said to be {\em piecewise linear} if there exists a partition $\br^n = P_1 \cup \ldots \cup P_N$ of the space $\br^n$ into finitely many polyhedral domains $P_k$ such that $q(\cdot)$ is linear on any of them: $q(x) = A_k x+b_k, \forall x \in P_k$.
\end{definition}
In this section, we consider the system
\begin{equation}
\label{eq.1a}
x_{t+1} = q[x_t], \qquad t=0,1,\ldots.
\end{equation}
under the following.
\begin{ass}
\label{ass.1}
The function $q(\cdot)$ is defined on $\br^n$, continuous, piecewise linear, and positively homogeneous
\begin{equation}
\label{hom}
q(\theta x) = \theta q(x), \qquad \forall x \in \br^n, \theta \in [0, \infty).
\end{equation}
\end{ass}
Then $q(0) =0$, and so $0$ is an equilibrium of \eqref{eq.1a}. Also in Definition~\ref{def.1}, the partition can be chosen so that any $P_k$ is a polyhedral cone (with the vertex at the origin) and $b_k=0$. Such a partition is said to be {\it conical} and $q$-{\it related}.

\begin{definition}
\label{def:differential}
Let $f : O \to \br^n$ be defined on an open subset $O \subset \br^n$ and let $x_\ast \in O$ be given. A continuous, piecewise linear, and positively homogeneous function $q : \br^n \to \br^n$ is called the {\em differential of} $f(\cdot)$ at the point $x_\ast$ (and denoted by $f^\prime(x_\ast)$) if
\begin{equation}
\label{defdiff}
\frac{f(x_\ast+h) - f(x_\ast) - q(h)}{\|h\|} \to 0 \qquad \text{\rm as} \quad h \to 0.
\end{equation}
\end{definition}
It is easy to check that the differential is unique (if exists) and also that the differential in the conventional sense is a particular case of the differential in the sense of Definition~\ref{def:differential}. Furthermore, \eqref{defdiff} holds if and only if there exists a function $\alpha : \br^n \to \br^n$ such that
\begin{equation}
\label{dec.first}
\begin{aligned}
  & 
  f(x) = f(x_\ast) + q(x-x_\ast) + \|x - x_\ast\| \alpha(x-x_\ast),
  \\
  & \alpha(h) \to 0 \quad \text{\rm as} \quad h \to 0.
\end{aligned}
\end{equation}
Finally, the definition of the differential is invariant with respect to the choice of the norm in $\br^n$.

\begin{prop}
\label{prop.diff}
Suppose that a function $f: O \subset \br^n \to \br^n$ is piecewise smooth in a neighborhood of a point $x_\ast \in O$. Then this function has a differential $f^\prime(x_\ast)$ in the sense of 
Definition~{{\rm \ref{def:differential}}}.
\end{prop}
\par
The proof of this proposition is prefaced by detailed description of this differential. To this end, we use the hypersurfaces from Definition~\ref{def:piecewise-smooth-function}. They  can be given by equations of the form $S_i = \{x \in \mathcal{O} : g_i(x) =0 \}$ provided that the neighborhood $\mathcal{O}$ is properly shrunk and a smooth function $g_i: \mathcal{O} \to \br$ is properly chosen for any $i$. Moreover, this can be accomplished so that the gradient $\nabla g_i(x) \neq 0 \; \forall x \in \mathcal{O}$. Now we put
$$
I:= \{i=1,\ldots,k: x_\ast \in S_i\}
$$
and define $B$ as the set of all maps $\mathfrak{b}=\{b_i\}_{i \in I}$ that assume only two values $1$ and $-1$. Given $\mathfrak{b} \in B$, we introduce the following sets
\begin{equation}
\label{def.sets}
\begin{aligned}
  & 
\mathfrak{C}_{\mathfrak{b}} := \{x \in \mathcal{O}: b_i g_i(x) <0 \; \forall i \in I\},
\\
&
K_{\mathfrak{b}} := \{h \in \br^n: b_i h^\top \nabla g_i(x_\ast) \leq 0 \; \forall i \in I\}.
\end{aligned}
\end{equation}
Since $\nabla g_i(x_\ast)$ is the normal to $S_i$ at point $x_\ast$, i) of Definition~\ref{def:non-tangential} implies that the both sets are non-empty, $x_\ast \in \overline{\mathfrak{C}_{\mathfrak{b}}}$, and $K_{\mathfrak{b}}$ is a polyhedral cone with the vertex at the origin.
Also, 
$\{\mathfrak{C}_{\mathfrak{b}}\}_{\mathfrak{b} \in B}$ is the variety of all connected components of the set $\mathcal{O} \setminus \big[ S_1 \cup \ldots \cup S_k\big]$ under the condition that $\mathcal{O}$ is properly shrunk; we assume that the last is accomplished. By ii) in Definition~\ref{def:piecewise-smooth-function}, $f(\cdot)$ can be extended from $\mathfrak{C}_{\mathfrak{b}}$ to a smooth function $f_{\mathfrak{b}}(\cdot)$ defined in an open neighborhood $O_{\mathfrak{b}}$ of $\overline{\mathfrak{C}_{\mathfrak{b}}}$. Finally, we define a map $q: \br^n \to \br^n$ by putting
\begin{equation}
\label{diff.pw}
q(h) := f^\prime_{\mathfrak{b}}(x_\ast) h \qquad \text{whenever} \quad h \in K_{\mathfrak{b}} \; \text{for some} \; \mathfrak{b} \in B,
\end{equation}
where the r.h.s. uses the ordinary Jacobian matrix $f^\prime_{\mathfrak{b}}(x_\ast)$.
\begin{lemma}
\label{lem.wd}
The map $q(\cdot)$ is well-defined.
\end{lemma}
\noindent{\bf Proof.} Since $\bigcup_{\mathfrak{b} \in B} K_{\mathfrak{b} \in B} = \br^n$ thanks to i) in Definition~\ref{def:non-tangential}, it suffices to show that
\begin{equation}
\label{entail}
h \in K_{\mathfrak{b}^\dagger} \cap K_{\mathfrak{b}^\lozenge} \Rightarrow 
f^\prime_{\mathfrak{b}^\dagger}(x_\ast) h = f^\prime_{\mathfrak{b}^\lozenge}(x_\ast) h.
\end{equation}
Let $h$ meet the premises of this entailment. Then
\begin{equation}
\label{def.ih}
I_h := \{i \in I: h^\top \nabla g_i(x_\ast) =0 \} \supset \{i \in I: b^\dagger_i \neq b^\lozenge_i\}.
\end{equation}
By invoking i) in Definition~\ref{def:non-tangential} once more, we infer that $\mathcal{S} := \{x \in \mathcal{O} : g_i(x) =0 \; \forall i \in I_h\}$ is a smooth manifold in a vicinity of $x_\ast$ and also that $h$ is tangential to $\mathcal{S}$ at $x_\ast$. Hence there exists a parametric curve $\gamma(\theta), \theta \in [0,\delta], \delta>0$ such that
$$
\mathcal{S} \ni \gamma(\theta) = x_\ast + \theta h + \oms(\theta) .
$$
So $g_i[\gamma(\cdot)] \equiv 0 \; \forall i \in I_h$. If $i \in I \setminus I_h$, \eqref{def.sets} and \eqref{def.ih} yield that $b^\dagger_i = b^\lozenge_i$ and $b^\dagger_i h^\top \nabla g_i(x_\ast) < 0$. It follows that $b^\dagger_i g_i[\gamma(\theta)] <0$ and $b^\lozenge_i g_i[\gamma(\theta)] <0$ for $\theta \approx 0$ and $i \in I \setminus I_h$. Overall, for $\theta \approx 0$, we have $b^\dagger_i g_i[\gamma(\theta)]  \leq 0$ and $b^\lozenge_i g_i[\gamma(\theta)] \leq 0$ and so $\gamma(\theta) \in \overline{\mathfrak{C}_{\mathfrak{b}^\dagger}} \subset O_{\mathfrak{b}^\dagger}, \gamma(\theta) \in \overline{\mathfrak{C}_{\mathfrak{b}^\lozenge}} \subset O_{\mathfrak{b}^\lozenge}$. By the continuity argument
\begin{equation}
\label{closure}
f(x) = f_{\mathfrak{b}}(x) \; \forall x \in \mathfrak{C}_{\mathfrak{b}} \Rightarrow f(x) = f_{\mathfrak{b}}(x) \; \forall x \in \overline{\mathfrak{C}_{\mathfrak{b}}}. 
\end{equation}
Thus we see that
\begin{equation}
\!\!\!\!\!\begin{aligned}
  & 
f[\gamma(\theta)] \!-\! f[x_\ast] \!=\! f_{\mathfrak{b}^\dagger}[\gamma(\theta)] \!-\! f_{\mathfrak{b}^\dagger}[x_\ast] \!=\! \theta f^\prime_{\mathfrak{b}^\dagger}(x_\ast) h \!+\! \oms(\theta), 
\\
&
f[\gamma(\theta)]\!-\!f[x_\ast]\!=\!f_{\mathfrak{b}^\lozenge}[\gamma(\theta)]\!-\!f_{\mathfrak{b}^\lozenge}[x_\ast] \!=\! \theta f^\prime_{\mathfrak{b}^\lozenge}(x_\ast) h \!+\! \oms(\theta)
\end{aligned}
\end{equation}
Therefore \eqref{entail} is valid. $\Box$
\begin{lemma}
\label{lemm.ll}
Any function $q(\cdot)$ satisfying Assumption~{\rm \ref{ass.1}} is globally Lipschitz continuous.
\end{lemma}
\par
The proof is similar to lemma~\ref{lemma l1}.

\begin{lemma}
\label{lem.last}
The map $q(\cdot)$ given by \eqref{diff.pw} is the differential of the map $f(\cdot)$ in the sense of Definition~{\rm \ref{def.1}}.
\end{lemma}
\noindent{\bf Proof.} 
Suppose the contrary. 
By \eqref{def.sets}, \eqref{diff.pw}, and Lemma~\ref{lem.wd}, the map $q(\cdot)$ satisfies Assumption~\ref{ass.1}.
So by Definition~\ref{def:differential}, \eqref{defdiff} fails to be true: there exists $\delta>0$, an infinite sequence $\{h_j\} \subset S_0^1$ of unit vectors $\|h_j\|=1$ and a sequence $\{\theta_j\} \subset (0,\infty)$ such that
\begin{equation}
\label{sontrr}
\begin{aligned}
  & 
  A_j:=\frac{\left\|f(x_\ast+\theta_j h_j) - f(x_\ast) - \theta_j q(h_j)\right\|}{\theta_j} \geq \delta, 
  \\
  &
   \theta_j \to 0 \quad \text{as} \; j \to \infty.
\end{aligned}
\end{equation}
By passing to a proper subsequence, we can also ensure that there exist $h_\infty \in S_0^1$ and $\mathfrak{b} \in B$ such that
$$
h_j \to h_\infty \quad \text{as} \; j  \to \infty, \qquad \text{and} \quad b_i g_i[x_\ast+\theta_j h_j] \leq 0 \; \forall i \in I.
$$
These and \eqref{def.sets} imply that $h_\infty \in K_{\mathfrak{b}}$ and so $q(h_\infty) = f^\prime_{\mathfrak{b}}(x_\ast) h_\infty $ by \eqref{diff.pw}, whereas $x_\ast+ \theta_j h_j \in \overline{\mathfrak{C}_{\mathfrak{b}}} \; \forall j$. Whence $f[x_\ast+ \theta_j h_j] = f_{\mathfrak{b}}[x_\ast+ \theta_j h_j] \; \forall j$ by \eqref{closure}. So
\begin{gather*}
A_j\!=\!
\frac{\left\|f_{\mathfrak{b}}(x_\ast\!+\!\theta_j h_j)\!-\!f_{\mathfrak{b}}(x_\ast)\!-\!\theta_j q(h_\infty)\!+\!\theta_j [q(h_\infty)\!-\!q(h_j)]\right\|}{\theta_j}
\\
\!\leq\! 
\frac{\left\|f_{\mathfrak{b}}(x_\ast+\theta_j h_j) - 
f_{\mathfrak{b}}(x_\ast)\!-\!\theta_j f^\prime_{\mathfrak{b}}(x_\ast) h_\infty \right\|}{\theta_j}
+ \left\| q(h_\infty)-q(h_j) \right\|
\\
\overset{\text{Lem.~\ref{lemm.ll}}}{\leq} \frac{\left\|f_{\mathfrak{b}}(x_\ast+\theta_j h_j) - f_{\mathfrak{b}}(x_\ast) - \theta_j f^\prime_{\mathfrak{b}}(x_\ast) h_\infty \right\|}{\theta_j}
 \!+\!c \left\|h_\infty-h_j \right\| ,
\end{gather*}
which tends to zero as $j \to \infty$
in violation of \eqref{sontrr}. This contradiction completes the proof. \hfill $\Box$

\noindent{\bf Proof of Proposition~\ref{prop.diff}} is immediate from Lemma~\ref{lem.last}. \hfill $\Box$

\subsection{Stability of the steady state for the piecewise-smooth function}
\label{app:stability}
\begin{prop}
\label{prop.1}
Suppose that Assumption~{\rm \ref{ass.1}} holds and the following claims are true:
\begin{enumerate}
\item[i)] The origin $0$ is the locally asymptotically stable equilibrium of the system \eqref{eq.1a};
\item[ii)] There exists a positively definite matrix $P = P^{\top} \in \br^{n \times n}$ and a conical $q$-related partition such that
\begin{equation}
\label{l.ineq}
A_k^\top P A_k \leq P \qquad \forall k .
\end{equation}
\end{enumerate}
Then the origin is a globally asymptotically stable equilibrium of 
\eqref{eq.1a} 
and there exists a natural $m$ and real $\eta \in (0,1)$ such that \begin{equation}
\label{lyap1}
V[q^m(x)] \leq \eta V[x] \quad \forall x \in \br^n, \qquad \text{\rm where} \quad V(x) := x^\top P x
\end{equation}
and $q^m := q \circ \cdots \circ q$ is the $m$th iteration of the map $q$.
\end{prop}
\par
The remainder of the section is devoted to the proof of Proposition~\ref{prop.1}, and so we posit that its assumptions are true.
We also introduce the $P$-related Euclidean norm $\|x\|_P := \sqrt{x^\top P x}$.
\begin{lemma}\label{lemma l1}
The function $q(\cdot)$ is Lipschitz continuous and its Lipschitz constant with respect to $\|\cdot\|_P$ equals $1$:
\begin{equation}
\label{lipsch}
\|q(x_2) - q(x_1)\|_P \leq \|x_2 - x_1\|_P \qquad \forall x_i \in \br^n.
\end{equation}
\end{lemma}
\noindent{\bf Proof.} Whenever $x_1,x_2$ lie in a common domain $P_k$, \eqref{lipsch} is straightforward from \eqref{l.ineq}.
By the continuity argument, this inequality extends on any $x_1,x_2 \in \overline{P}_k$ from the closure $\overline{P}_k$ of $P_k$.
For an arbitrary pair $x_1, x_2 \in \br^n$, there exists a finite sequence $0= \theta_0 < \theta_1 < \ldots < \theta_s < \theta_{s+1} = 1$ such that
$x(\theta_i)$ and $x(\theta_{i+1})$ lie in a common domain $\overline{P}_{k(i)}$ for any $i=0, \ldots,s$, where $x(\theta) := (1-\theta)x_1 + \theta x_2$. 
Since
\begin{equation}
\begin{aligned}
  & 
    \|q[x(\theta_{i+1})] - q[x(\theta_i)]\|_P 
    \leq 
    \|x(\theta_{i+1}) - x(\theta_i)\|_P, 
  \\
  &
   \forall i=0, \ldots,s,
\end{aligned}
\end{equation}
Then
\begin{equation}
\begin{aligned}
  & 
    \|q[x_2] - q[x_1]\|_P 
    \leq \sum_{i=0}^s \|x(\theta_{i+1}) - x(\theta_i)\|_P
    \\
    & 
   \leq \|x_2-x_1\|_P. 
  \qquad \Box
\end{aligned}
\end{equation}
\par
It follows that for any natural $m$,
\begin{equation}
\label{lipsch1}
\|q^m(x_2) - q^m(x_1)\|_P \leq \|x_2 - x_1\|_P \qquad \forall x_i \in \br^n.
\end{equation}
\begin{lemma}
\label{lem.2}
There exists a natural $m$ and $\eta \in (0,1)$ such that \eqref{lyap1} holds.
\end{lemma}
\noindent{\bf Proof.}
Since $x_t(\theta a) = \theta x_t(a)$ for all $t=0,1,\ldots$ and $a \in \br^n$ by \eqref{hom}, assumption i) of Proposition~\ref{prop.1} implies that
\begin{equation}
\label{converge}
x_t(a) \to 0 \quad \text{as} \quad t \to \infty \qquad \forall a \in \br^n.
\end{equation}
Now we pick $\eta \in (0,1)$ and a finite $\sqrt{\eta}/2$-net $a_1, \ldots, a_K \in S_0^1$ in the unit sphere $S_0^1 := \{x \in \br^n: \|x\|_P=1\}$:
$ x \in S_0^1 \Rightarrow \exists j = j(x)=1,\ldots, K : \|x - a_j\|_P < \sqrt{\eta}/2.$
Such a net does exist since the sphere is compact. Now we apply \eqref{converge} to $a:= a_j$ and note that the convergence is uniform over the finite variety of $j$'s. So there exists a natural $m$ such that
$$
\|q^t(a_j)\|_P = \|x_t(a_j)\|_P < \sqrt{\eta}/2 \qquad \forall t \geq m, j=1, \ldots, K.
$$
Hence for $t \geq m$, we have
\begin{equation}
\begin{aligned}
  & x \in S_0^1 \Rightarrow 
  \\
  & \|q^t(x)\|_P \leq \|q^t(a_{j(x)})\|_P + \|q^t(a_{j(x)}) - q^t(x)\|_P 
  \\
  &
   \overset{\text{\eqref{lipsch1}}}{<} \sqrt{\eta}/2 + \|a_{j(x)} - x\|_P \leq \sqrt{\eta}/2+\sqrt{\eta}/2 = \sqrt{\eta}.
\end{aligned}
\end{equation}
\par
For $x=0$, \eqref{lyap1} is clear. Let $x \neq 0$. Then $x/\|x\|_P \in S_0^1$ and so
\begin{gather*}
\sqrt{\eta} \geq \left\|q^t\left( \frac{x}{\|x\|_P} \right)\right\|_P = \frac{\|q^t(x)\|}{\|x\|_P} \Rightarrow \|q^t(x)\| \leq \sqrt{\eta} \|x\|_P,
\\
V[q^t(x)] = \|q^t(x)\|^2 \leq \eta \|x\|_P^2 = V(x).
\end{gather*}
\noindent{\bf Proof of Proposition~\ref{prop.1}.} Whereas \eqref{lyap1} holds by Lemma~\ref{lem.2}, the first claim of the proposition follows from \eqref{lyap1}. \hfill $\Box$

\begin{theorem}
\label{thm.main}
Suppose that the following claims hold:
\begin{itemize}
\item A point $x_\ast \in O$ is an equilibrium for the discrete-time dynamical system
\begin{equation}
\label{eq.1}
x_{t+1} = f[x_t], \quad x_t \in \br^n, \qquad t=0,1,\ldots.
\end{equation}
\item There exists a differential $q(\cdot) = f^\prime(x_\ast)$ (in the sense of Definition~{\rm \ref{def:differential}}) of $f(\cdot)$ at $x_\ast$;
\item The properties {\rm i)} and {\rm ii)} from Proposition~{\rm \ref{prop.1}} hold for this differential.
\end{itemize}
This theorem reduces analysis of stability of the primal nonlinear system to stability analysis of a simpler nonlinear system, which is piece-wise linear and positively homogeneous, unlike the primal one. We refer the reader to 
\cite{feng2002stability,groff2019stability,iervolino2015cone,liu2014stability,sun2010stability,waitman:tel-01917511}
for surveys of tools available for stability analysis of piecewise linear systems.
Then the equilibrium $x_\ast$ of the system \eqref{eq.1} is locally exponentially asymptotically stable.
\end{theorem}
\par
The proof of this theorem is prefaced by the following.
\begin{lemma}
\label{lem.dif}
For any natural $m$, the iteration $q^m(\cdot)$ is the differential of $f^m(\cdot)$.
\end{lemma}
\noindent{\bf Proof.}
We invoke the matrix $P$ from ii) in Proposition~{\rm \ref{prop.1}} and the $P$-related norm $\|\cdot\|_P$.
The proof will be via induction on $m$. For $m=1$, the claim is evident.
Let it be true for some $m$, i.e., let
\begin{equation}
\begin{aligned}
  & f^m(x) = \underbrace{f^m(x_\ast)}_{= x_\ast} + \Delta^m_x,
\end{aligned}
\end{equation}
where
\begin{equation}
\begin{aligned}
  & \Delta^m_x:= q^m(x-x_\ast) + \|x - x_\ast\|_P \alpha_m(x-x_\ast)
\end{aligned}
\end{equation}
and 
\begin{equation}
\label{m.diff}
\begin{aligned}
  & \quad\alpha_m(h) \to 0 \quad \text{\rm as} \quad h \to 0.
\end{aligned}
\end{equation}
The rest of the proof follows from the fact that $q^m(\cdot)$ is continuous and positively homogeneous
\begin{equation}
\begin{aligned}
  & \frac{\|f^{m+1}(x_\ast+h) - f(x_\ast) - q^{m+1}(h)\|_P}{\|h\|_P} 
  \\
  &
   \leq  \|\alpha_m(h)\|_P+ \big[c_m+\|\alpha_m(h)\|_P \big] \|\alpha[\Delta^m_{x_\ast+h}]\|_P \to 0
\end{aligned}
\end{equation}
as $h \to 0$, which completes the proof. \hfill $\Box$

\noindent{\bf Proof of Theorem~\ref{thm.main}.} We invoke $P, V(\cdot), m$, and $\eta$ from Proposition~\ref{prop.1}, and observe that
\begin{equation}
\begin{aligned}
  & V[f^m(x) - x_\ast] \overset{\text{\eqref{m.diff}}}{=} V\big[ q^m(x-x_\ast) + \|x - x_\ast\|_P \alpha_m(x-x_\ast) \big]
  \\
  &
 \overset{\text{\eqref{lyap1}}}{\leq} 
 V(x-x_\ast) \mathcal{K},
 \\
 & \quad \mathcal{K} = \Big\{ \eta + V \big[ \alpha_m(x-x_\ast) \big]
+ 2 c_m \|\alpha_m(x-x_\ast)\|_P \Big\}.
\end{aligned}
\end{equation}
Now we pick $\overline{\eta} \in (\eta,1)$. By the continuity argument, there exists $\delta>0$ such that whenever $\|x-x_\ast\|_P < \delta$, we have  $\mathcal{K} < \overline{\eta}$ and so $V[f^m(x) - x_\ast] \leq \overline{\eta} V(x-x_\ast)$. Thus, we see that $x \mapsto V(x-x_\ast)$ is a strong Lyapunov function for the discrete-time dynamical system
\begin{equation}
\label{traj}
x_{t+1} = f^m(x_t), \qquad t=0,1,\ldots.
\end{equation}
By the second Lyapunov principle, there exist $\varkappa >0$ and $c>0$ such that
$$
\|x^m_t(a) - x_\ast\| \leq c \overline{\eta}^{t/2} \|a\| \; \forall t \geq 0 \quad \text{whenever} \; \|a\| < \varkappa,
$$
where $\{ x^m_t(a)\}_t$ is the trajectory of \eqref{traj} starting with $x_0^m=a$.
\par
By Lemma~\ref{lem.dif}, any iteration $f^k(\cdot)$ has a differential at the point $x_\ast$. Hence \eqref{m.diff} 
imply existence of $\varepsilon_k>0$ and $c_k>0$ such that
$$
\|f^k(x) - x_\ast\| \leq c_k \|x - x_\ast\| \quad \text{whenever} \quad \|x-x_\ast\| < \varepsilon_k.
$$
Now we put $c_0:= 1$ and
$\varepsilon := \min \left\{ \min_{k=1,\ldots,m} \varepsilon_k; \min_{k=0,\ldots,m} \frac{\varkappa}{c_k} \right\}, \quad C:= c \max_{k=0,\ldots,m} c_k.$
Then whenever $\|a - x_\ast\| < \varepsilon$, we have $\|f^k(a) - x_\ast \| < \varkappa \; \forall k=0, \ldots, m$ and so
\begin{equation}
\begin{aligned}
  & \|x^m_\tau[f^k(a)] - x_\ast\| \leq c \overline{\eta}^{\tau/2} \|f^k(a) - x_\ast\| 
  \\
  &
  \leq  c c_k \overline{\eta}^{\tau/2} \|a-x_\ast\|
= C \overline{\eta}^{\tau/2} \|a-x_\ast\|\; 
  \\
  & \forall \tau \geq 0, k=0,\ldots,m.
\end{aligned}
\end{equation}
Let $x_t(a)$ stand for the trajectory of \eqref{eq.1} that starts with $x_0(a) =a$ and $\lfloor \cdot \rfloor$ for the integer floor. Then $x_t(a) = x^m_{\lfloor t/m \rfloor}[f^{t - m\lfloor t/m \rfloor}(a)]$ and so
\begin{equation}
\begin{aligned}
  & \|x_t(a) -x_\ast\| = \left\| x^m_{\lfloor t/m \rfloor}[f^{t - m\lfloor t/m \rfloor}(a)] - x_\ast\right\| 
  \\
  &
 \leq \frac{C}{\sqrt{\overline{\eta}}} (\sqrt{\overline{\eta}})^{t/m} \|a-x_\ast\| \leq C_\ast \eta_\ast^t \|a-x_\ast\|,
\end{aligned}
\end{equation}
where $C_\ast:= \frac{C}{\sqrt{\overline{\eta}}}$ and $\eta_\ast := (\overline{\eta})^{1/(2m)} \in (0,1)$. This completes the proof. \hfill $\Box$

\section{Proof of theorem~\ref{thm.holdin}}
\label{app:theorem-proof}
Stability analysis of the zero steady state of model \eqref{eq:complete-model-ab} is studied by applying
theorem~\ref{thm.main} from 
in Appendix~\ref{app:stability-theory}.
In subsection~\ref{ssec:piecewise-smoothness} 
we show that the right-hand side of \eqref{eq:complete-model-ab} is piecewise smooth.
In subsection~\ref{ssec:differential} system \eqref{eq:complete-model-ab} is linearized (resulting in piecewise-linear map).
In subsection~\ref{ssec:stability-of-linear-system} stability of the origin for linearized  system is studied.
In subsection~\ref{ssec:stability-theorem} stability of the linearized system is used to prove stability of the origin for non-linear system.
\label{sec:local-stab-proof}
\subsection{Piecewise-smoothness of the right-hand side}
\label{ssec:piecewise-smoothness}
It is easy to see, that the right-hand side of \eqref{eq:complete-model-ab} $f:\mathcal{O}\subset \br^2\to\br^2$ satisfies the conditions 1)--3) for piecewise smooth function in a small neighborhood $\mathcal{O}$ of the origin
(see Definition~\ref{def:piecewise-smooth-function}, Appendix~\ref{app:stability-theory}):
  \begin{enumerate}
    \item The function $f(\cdot)$ is continuous in a small neighborhood $\mathcal{O}$ of equilirbium $(0,0)$;  
    \item From \eqref{eq:complete-model-ab} in a small neighborhood of equilibrium $\mathcal{O}$ there are three curves $S_1$, $S_2$, $S_3$
    dividing the phase space:
    \begin{itemize}
  \item $S_1$: $p_k=0$;
  \item $S_2$: $p_k> 0$, $c_k=0$;
  $u_k = \frac{p_k}{1-p_k}$ for $0<p_k<1$;
  \item $S_3$: $p_k<0$, $l_k=1$.
\end{itemize}
     The function $f(\cdot)$ is continuously differentiable on any connected component $\mathfrak{C}$ of the set $\mathcal{O} \setminus \big[ S_1 \cup S_2 \cup S_3\big]$ and, moreover, can be extended from $\mathfrak{C}$ to an open neighborhood of the closure $\overline{\mathfrak{C}}$ with retaining this smoothness;
        \item The curves $S_1$, $S_2$, and $S_3$ are mutually non-tangential on $\mathcal{O}$.
    \end{enumerate}

\subsection{Linearization}
\label{ssec:differential}
One can compute the differential $q(\cdot)$ of $f(\cdot)$ 
(see Definition~\ref{def:differential} and Proposition~\ref{prop.diff}, Appendix~\ref{app:stability-theory}) as follows: 
\begin{equation}
  \label{eq:linearized}
  \begin{aligned}
    & x_{k+1}=q(x_k)=A_j x_k,\ x_k\in R_j,\ j=1,2,3,4,\\
      & A_1 = \frac{1}{1+\alpha}
    \begin{bmatrix}
      1  & -1\\
      2\beta & 1+\alpha-2\beta
    \end{bmatrix},
    \\ 
    & A_2 = 
    \begin{bmatrix}
      1-\alpha  & -1\\
      2\beta(1-\alpha) & 1-2\beta
    \end{bmatrix},\\
    & A_3 = 
    \begin{bmatrix}
      1  & -1\\
      2\beta & 1-2\beta
    \end{bmatrix},
    \\
    & A_4 = \frac{1}{\alpha+1}
    \begin{bmatrix}
      1-\alpha  & -1\\
      2\beta(1-\alpha) & 1+\alpha-2\beta
    \end{bmatrix}.
  \end{aligned}
\end{equation}
where the lines $L_1$, $L_2$, $L_3$ separating conical regions $P_j$
are defined by equations conical partitioning:
\begin{itemize}
  \item $L_1$: $p_k=0$;
  \item $L_2$: $p_k\geq0$, $u_k=p_k$;
  \item $L_3$: $p_k\leq0$, $u_k=(1-\alpha)p_k$.
\end{itemize}
\subsection{Stability of the linearized system}
\label{ssec:stability-of-linear-system}
Consider linearized system \eqref{eq:linearized} and the quadratic Lyapunov function
  \begin{equation}
    \begin{aligned}
      & V(x)=x^T Hx,\quad
      H = \begin{bmatrix}
        2\beta & -\beta\\
        -\beta & 1
      \end{bmatrix}.
    \end{aligned}
  \end{equation}
One can check that $V(x)$ is positive definite for $0<\beta<2$  and is non-increasing along the trajectories of the linearized system:
\begin{itemize}
     \item $x\!\in\!P_1$:
        $V(q(x))-V(x)
                \!=\! -x^T
                \frac{2\beta\alpha}{(1+\alpha)^2}
                \!\begin{pmatrix}
                1 & -1\\
                -1 & 1
              \end{pmatrix}\!x\leq0$,
     \item $x\!\in\!P_2$:    
        $V(q(x))-V(x)
                = -x^T
                \begin{pmatrix}
                2\beta \alpha & 0\\
                0 & 0
              \end{pmatrix}x\leq0$,    
  \item $x\in P_3$:
  $V(q(x))-V(x)=0$,
  \item $x\!\in\! P_4$:
      $V(q(x))-V(x)\!=\!-x^T\frac{2\alpha\beta}{(1+a)^2}
                \!\begin{pmatrix}
                  4 & -2\\
                  -2 & 1
                \end{pmatrix}\!x\leq0$.    
\end{itemize}
Thus, the origin of the linearized system is stable.
The exponential stability of the origin for $0<\beta<2$, $\beta\neq \frac{3}{2}$ 
can be proved using discrete-time analog of the LaSalle invariance principle (see, e.g. \cite{Bof2018}).
First, we define a set 
\begin{equation}
  \begin{aligned}
    & E=\{x|V(x)=V(q(x))\}.
  \end{aligned}
\end{equation}
and show that $V(x)=V(q(x))$ for $x_k\in E= P_3\cup y_1$, where $y_1$ is a ray
\begin{equation}
  \label{eq:ray-y1}
  \begin{aligned}
    & y_1 = \{x|x\in r[-1,-2]^T,\quad r\geq 0\}.
  \end{aligned}
\end{equation}
Then we seek for the largest positively invariant set $M\subset E$. For $0<\beta<2$, $\beta\neq \frac{3}{2}$ one obtains $M=\{0\}$ and for $\beta=\frac{3}{2}$ the set $M$ is comprised of three rays (the lower and upper boundaries of $S_3$ and a special ray $y_1$).
Thus, the origin is asymptotically stable for $0<\beta<1$, $\beta\neq \frac{3}{2}$ (all necessary details of the proof are provided in 
\cite{Orla-2013-review}).

\subsection{Local stability of the nonlinear system}
\label{ssec:stability-theorem}
Now we have the following: 
\begin{itemize}
  \item The origin is an equilibrium for the discrete-time dynamical system $x_{k+1} = f(x_k)$;
  \item There exists the differential $q(\cdot) = f^\prime(x_\ast)$ (in the sense of Definition~{\rm \ref{def:differential}}) of 
  the function $f(\cdot)$ at $(0,0)$;
  \item
  The linearized system is asymptotically stable  
  (see properties {\rm i)} and {\rm ii)} 
  from Proposition~{\rm \ref{prop.1}}, Appendix~\ref{app:stability-theory}).
\end{itemize}
Then, Theorem~\ref{thm.main} from Appendix~\ref{app:stability-theory}
yields that the origin of system \eqref{eq:complete-model-ab} is uniformly exponentially stable for $0<\beta<2$, $\beta\neq \frac{3}{2}$, $0<\alpha<1$.
Note that for $\beta=\frac{3}{2}$ there is infinite number of period-3 limit cycles in linearized system \eqref{eq:linearized}, while these limit cycles are non-existent in nonlinear system \eqref{eq:complete-model-ab}, which makes the origin locally stable for $\beta=\frac{3}{2}$.

\section{Instability for $\alpha>0$, $\beta>2$}
\label{app:instability}
Here we use Lemma 5 from \cite{Orla-2013-review} to show that there is a trajectory of linearized system \eqref{eq:linearized} diverging from the origin.

It is easy to see, that the eigenvalues of $A_4A_3$ are real, distinct, and larger eigenvalue $\lambda_1>1$ has an associated eigenvector $x_1$ in the right-hand canonical form with the slope 
$\mu=(\lambda_1(a+1)+(a+2\beta-1))/(a_2\beta-2)$.
Since $\mu>1$ we get $x_1\in P_3$. Therefore,
\begin{equation}
  \begin{aligned}
    & A_3x_1=(\mu-1)\begin{bmatrix}
      -1\\
      \frac{2\beta+\mu(1-2\beta)}{\mu-1}
    \end{bmatrix}.
  \end{aligned}
\end{equation}
Since $(2\beta+\mu(1-2\beta))/(\mu-1)<\alpha-1$, that is 
$\mu>(\lambda_1(a+1)+(a+2\beta-1))/(a_2\beta-2)$, we get $A_3x_1\in P_4$. Accordingly, trajectory starting at $x_1\in P_3$ goes in one step to $A_3x_1\in P_4$ and in one further step to $A_4A_3x_1=\lambda_1 x_1\in P_3$, because $\lambda_1$ is real and positive. As $\lambda_1>1$, it is clear that the resulting orbit diverges.

\end{appendices}


  \begin{IEEEbiography}
 [\vspace{-0.8cm}{\includegraphics[width=1in,height=1.25in,clip,keepaspectratio]{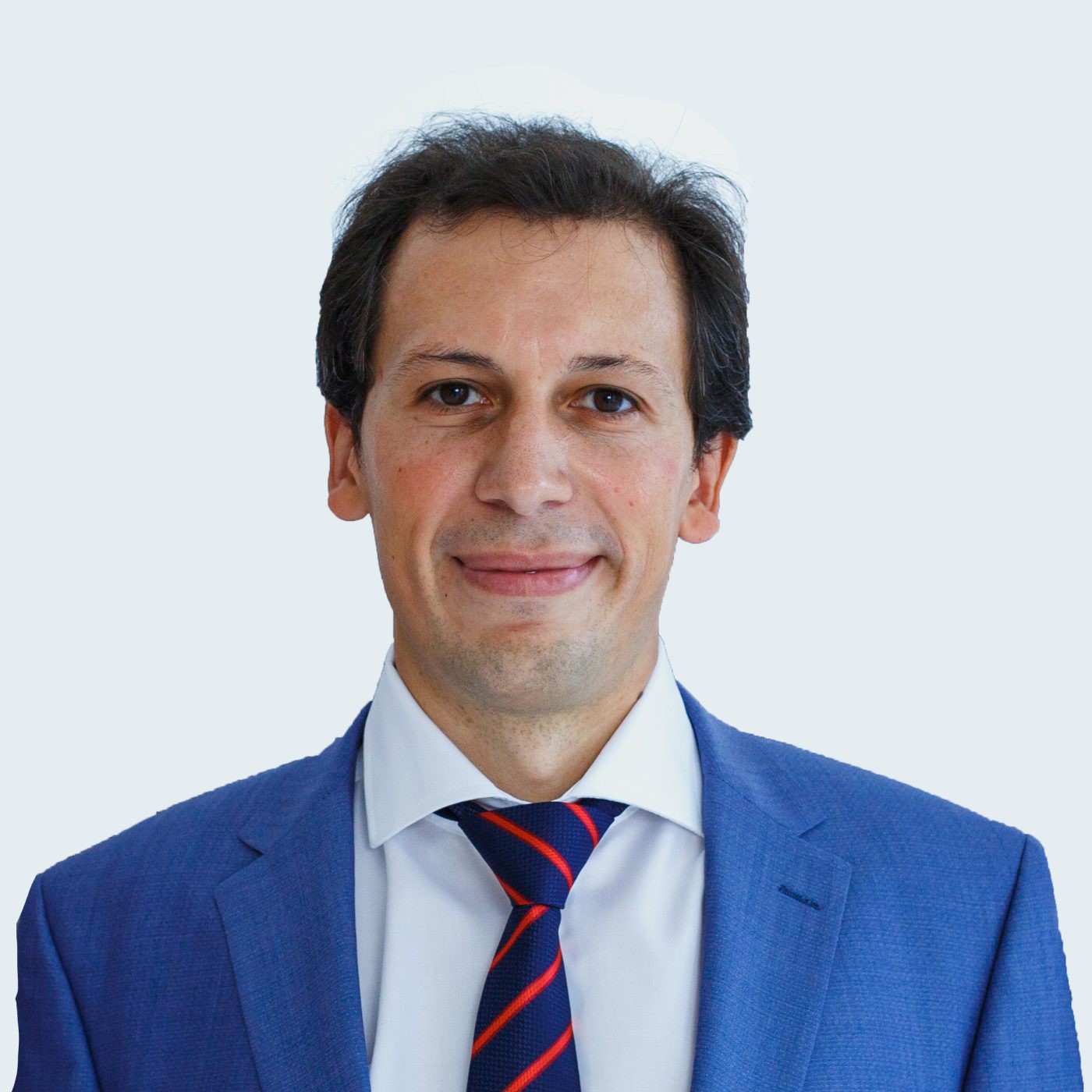}}]
 {Nikolay Kuznetsov}
  graduated from the St. Petersburg University in 2001. In 2004 he received the Candidate of Science degree and in 2016 the Doctor of Science degree from St. Petersburg University (Russia), where he is currently Professor and Head of the Department of Applied Cybernetics. In 2018 and 2020, the research group chaired by Prof. Kuznetsov was awarded the status of the Leading Scientific School (Center of Excellence) of Russia in the field of mathematics and mechanics. In 2020 he was named Professor of the Year in the field of mathematics and physics in Russia.
  In 2008, he defended his Ph.D. degree at the University of Jyv\"{a}skyl\"{a} (Finland),
  where now is Visiting Professor and co-chair of 
  the Finnish-Russian Educational \& Research program organized 
  together with St. Petersburg University. 
  In 2020 he was elected as foreign member of the Finnish Academy of Science and Letters.
  Since 2018, he is Head of the Laboratory of information and control systems at the Institute for Problems in Mechanical Engineering of the Russian Academy of Science.
   His research interests include nonlinear control systems,
 stability and oscillations in dynamical systems, theory of hidden oscillations,
 hidden attractors,
 phase-locked loop nonlinear analysis.    
\newline
 E-mail: nkuznetsov239@gmail.com (corresponding author)
\end{IEEEbiography}

\begin{IEEEbiography}
 [\vspace{-0.8cm}{\includegraphics[width=1in,height=1.25in,clip,keepaspectratio]{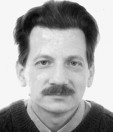}}]
{Alexey Matveev}
 was born in Leningrad, Russia, in 1954. He received the M.S. and Ph.D. degrees in applied mathematics and engineering cybernetics both from the Leningrad University, St. Petersburg, Russia, in 1976 and 1980, respectively.

He is currently a Professor in the Department of Mathematics and Mechanics, Saint Petersburg University. His research interests include control over communication networks, hybrid dynamical systems, and navigation and control of mobile robots.
  \newline
E-mail: almat1712@yahoo.com
\end{IEEEbiography}
\begin{IEEEbiography}
 [\vspace{-0.8cm}{\includegraphics[width=1in,height=1.25in,clip,keepaspectratio]{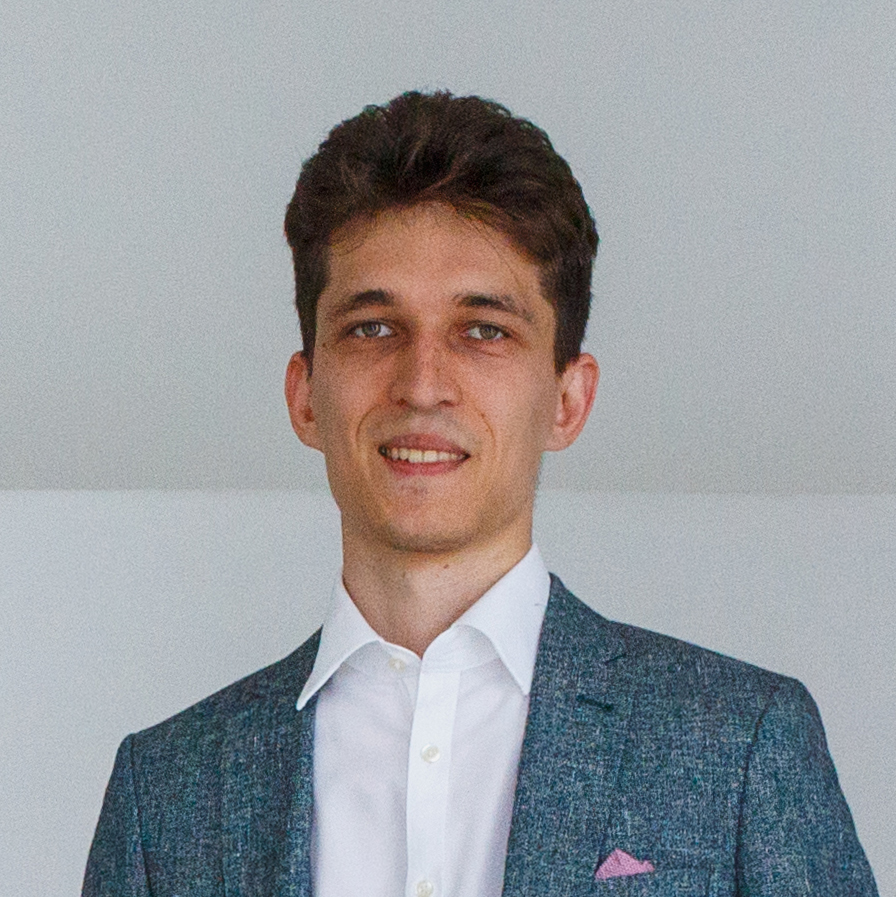}}]
{Marat Yuldashev}
received his Candidate degree from St.Petersburg State University, Russia
(2013) and PhD from the University of Jyv\"{a}skyl\"{a}, Finland (2013) in the framework of joint Russian-Finnish PhD program.
He is currently a Professor at Saint-Petersburg University.
His research interests cover nonlinear models of phase-locked loops and Costas loops, and SPICE simulation.
  \newline
E-mail: maratyv@gmail.com 
\end{IEEEbiography}
\begin{IEEEbiography}
 [\vspace{-0.8cm}{\includegraphics[width=1in,height=1.25in,clip,keepaspectratio]{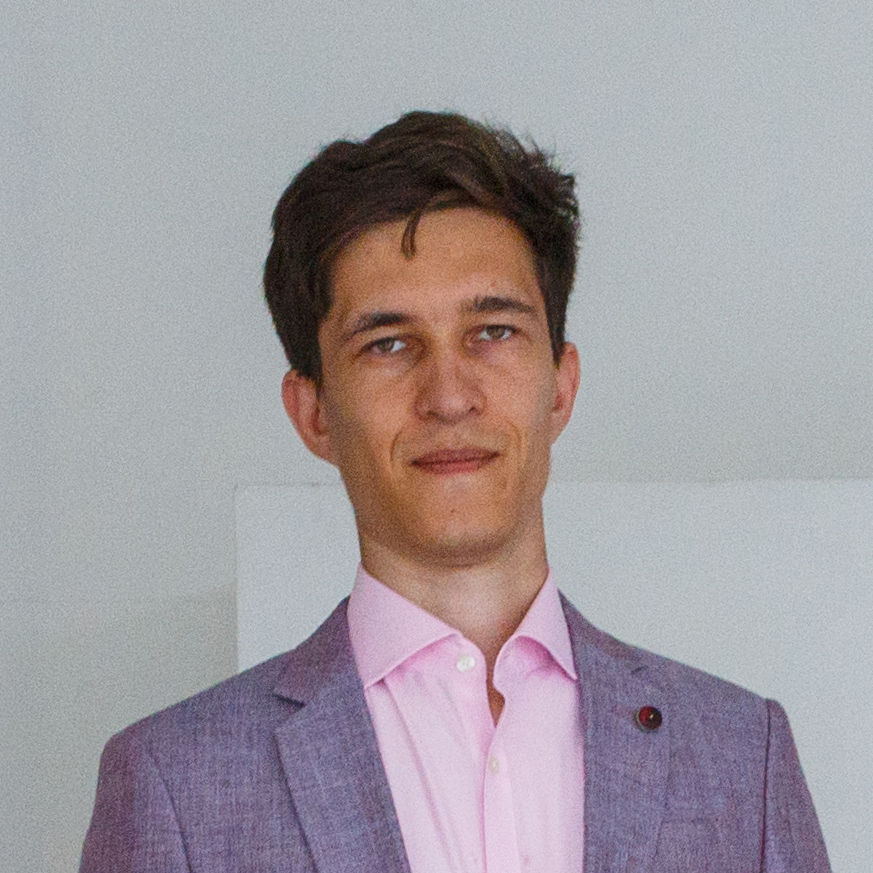}}]
{Renat Yuldashev}
received his Candidate degree from St.Petersburg State University, Russia
(2013) and PhD from the University of Jyv\"{a}skyl\"{a}, Finland (2013) in the framework of joint Russian-Finnish PhD program.
He is currently a Professor at Saint-Petersburg University.
His research interests cover synchronization of power converters in electrical grids, nonlinear analysis of PLLs.
  \newline
E-mail: renatyv@gmail.com 
\end{IEEEbiography}

\end{document}